\documentstyle[aps,preprint,psfig,eqsecnum]{revtex}
\tightenlines
\begin{document}
\draft
%\preprint{}
\title{Post-Newtonian gravitational radiation 
and equations of motion via direct
integration of the relaxed Einstein equations. \\
I. Foundations
}
\author{Michael E. Pati and Clifford M. Will }
\address{McDonnell Center for the Space Sciences,
Department of Physics, \\
Washington University, St. Louis, Missouri 63130}
\date{\today}
%\twocolumn[
\maketitle
%\widetext
\begin{abstract}
We present a self-contained framework called Direct Integration of the
Relaxed Einstein Equations (DIRE) for calculating equations of
motion and gravitational radiation emission for isolated gravitating
systems based on the post-Newtonian approximation.  We cast the Einstein 
equations into their ``relaxed'' form of a flat-spacetime wave
equation together with a harmonic gauge condition, and solve the equations
formally as a retarded integral over the past null cone of the 
field point (chosen to be within the near zone when calculating
equations of motion, and in the far zone when calculating gravitational 
radiation).  The ``inner'' part of this integral 
(within a sphere of radius $\cal
R \,\sim$ one gravitational wavelength) is approximated in a slow-motion 
expansion using standard techniques; the ``outer'' part, extending over the
radiation zone, is evaluated using a null integration variable.  
We show generally and explicitly that all contributions to the inner
integrals that depend on $\cal R$ cancel corresponding terms from the
outer integrals, and that the outer integrals converge at infinity,
subject only to reasonable assumptions about the past behavior of the source.
The method cures defects that plagued previous ``brute-force'' slow-motion
approaches to motion and  gravitational radiation for isolated systems.
We detail the procedure for iterating the solutions in a weak-field,
slow-motion approximation, and derive expressions for the
near-zone field through 3.5 post-Newtonian order in terms 
of Poisson-like potentials.
\end{abstract}
%\pacs{}
%]
%\narrowtext
\section{INTRODUCTION}
\label{sec:intro}

The motion of multiple, isolated bodies under their 
mutual gravitational attraction
and the resulting emission of gravitational radiation 
is a  long-standing problem
that dates back to the first years following the publication of
general relativity (GR).  It has at times been controversial (for a
thorough review see \cite{damour300}).
In 1916 Einstein calculated the gravitational
radiation emitted by a laboratory-scale object using the linearized
version of GR \cite{AE}.  Some of his assumptions were questionable and
his answer for the energy flux was off by a factor of two (an error
pointed out by Eddington \cite{eddington}).  In 1916, de Sitter
\cite{desitter} derived N-body equations of motion in what later would
be termed the post-Newtonian (PN) approximation.  
However, his equations contained an error that was
discovered in the course of a disputed claim by Levi Civita
\cite{levicivita} 
that the center of mass of a binary star
system would suffer a ``self-acceleration''.  Eddington and Clark
\cite{eddingtonclark}
corrected the error, and found no self-acceleration.
Einstein, Infeld and Hoffman \cite{EIH} 
attempted to demonstrate explicitly that
the Einstein equations alone imply equations of motion, by matching
solutions of the vacuum equations, expanded in a weak-field, slow
motion approximation, to fields representing the near-zone fields of
``point'' masses, working to first PN order.  
The result was the well-known EIH N-body equations
of motion.  Other highlights in this early history of the problem of motion
include the development of the post-Newtonian approximation for fluid
sytems by Fock \cite{fock} and Chandrasekhar
\cite{chandra}, its extension by
Chandrasekhar and later workers to 2.5PN order 
\cite{chandranutku,chandraesposito}, and the development of
equations of motion for spinning bodies by Papapetrou and
collaborators \cite{papapetrou}.  

Gravitational theory presents one problem essentially identical to
that of electromagnetic theory: how to mesh the
natural solution of the field equations in the near zone where the bodies
reside, which involves slow-motion expansions and instantaneous fields,
with the solution in the far zone, which involves retarded fields.  
Such a meshing is needed if one is to calculate the effects of
gravitational radiation reaction that result from the emission of
enegy and angular momentum to infinity.
One approach to resolving this problem was that
of matched asymptotic expansions.  Although well-rooted in applied
mathematics, it was first expounded in 1970 as a powerful technique for
electromagnetic and gravitational problems by Burke 
\cite{burke}.  Another, related approach is the
``post-Minkowskian'' framework, elaborated and developed most fully by
Blanchet and Damour and their collaborators 
\cite{bd86,bd88,bd89,di91,bdtail,luc95}.

A second important problem of gravitation, which distinguishes it from
electromagnetism, is the non-linearity of Einstein's equations.
Gravitation itself acts as a source of gravitation.  Consequently
this source
extends over all space, resulting in the
possibility of divergent or ill-defined integrals.
In many ways, this has been the most serious difficulty to overcome.
Techniques for resolving it have ranged from sweeping the difficulties
under the rug, to the sophisticated analytic regularization methods of the
post-Minkowskian program.  A central thrust of this paper is to
present a straightforward method for resolving this difficulty.

A third ``problem'', which is less a problem for gravitation than it
is for
electromagnetism, is that of ``point'' sources.  In electromagnetic
theory, where there is a belief that fundamental charges like the
electron are point-like, the singular nature of the fields at the
source has led to problems of mass regularization, especially in deriving
equations of electromagnetic radiation reaction; it also raises issues
of the boundary between classical and quantum electrodynamics.  In
gravitation theory, this is less of an issue of principle, because the
primary interest is in the motion of and radiation by astrophysical
systems, whose members are clearly not point masses.  Instead, the use
of ``point'', i.e. delta-function sources is meant as an efficient
means of approximating the mass distribution of bodies that are
nearly spherical and that are small compared to the typical
separation between them, so that tidal effects, which depend on the
finite size of the bodies, can be ignored.  Here the issue is how to
make use of a point mass approximation (which simplifies many
calculations) in a way that captures all the physics without
introducing spurious effects.

A fourth problem is of a technical nature: in electromagnetic theory,
radiation damping in the equations of motion occurs at order $(v/c)^3$
beyond the simple Coulomb forces between charges, and is relatively
easy to compute in a systematic approximation method, modulo the other
problems listed above.  By contrast, gravitational radiation damping
occurs at order $(v/c)^5$ beyond Newtonian gravity, and requires a
higher order of approximation that captures all relevant
contributions.  Over the years, numerous inequivalent
results have been quoted for the leading gravitational radiation
reaction effects.   One finds published papers in which the 
coefficient in the relevant formula has ranged from 
$-21/16$ 
to the correct coefficient of unity); a study by Walker and Will \cite{cranks}
showed that the divergent results were all the
simple consequence of missing one or more terms that contribute to the final
answer.

These four ``problems'' were the origin of the so-called ``quadrupole
controversy'', which arose from a critique by Ehlers and colleagues 
\cite{ehlers} of
the foundations of the quadrupole formula for the leading-order gravitational
radiation energy flux and orbital damping.  This critique had the
beneficial effect of spurring new research on those foundations, including
a study of the systematic structure of the approximation sequence of
Einstein's equations in a slow-motion, weak-field approach; analysis
of energy balance as an argument for connecting the far-zone energy
flux to the near-zone damping forces, and elaboration of the
post-Minkowskian approach, among others (see \cite{damour300} for a
review).  The work inspired by the Ehlers critique served to confirm
the quadrupole formula and to strengthen its foundations.  The
ultimate test, of course, came in 1979 with the announcement of the
measurement of orbital damping of the binary pulsar PSR 1913+16 in
agreement with the quadrupole formula \cite{taylor79}; 
current results agree to better
than 0.5 percent \cite{taylor94}.

The problem of motion and radiation has received renewed interest
since 1990, with the proposal for large-scale laser interferometric
gravitational-wave observatories, such as the LIGO project in the US, 
and the realization that a leading
candidate source of detectable waves would be the radiation-reaction
driven inspiral of a binary system of compact objects (neutron stars
or black holes) \cite{snowmass}.  Furthermore, it was noted \cite{3min} that the leading method
for data analysis of signals from such systems, optimal matched
filtering, would require theoretical template waveforms that are
accurate (primarily in the evolution of the orbital frequency or
phase) well beyond the leading-order prediction of the
quadrupole formula, possibly as high as corrections of order
$(v/c)^6$.  
 
This presented a major theoretical challenge: to calculate the motion
and radiation to very high PN order, a formidable algebraic task,
while addressing each of the problems listed above sufficiently well
to ensure that the results were physically meaningful.  This challenge
was taken up by three groups of workers.

One group, headed by Blanchet, Damour and Iyer 
\cite{bd86,bd88,bd89,di91,bdtail,luc95}, used the
post-Minkowskian (PM) approach to derive the gravitational waveform,
equations of motion 
and energy flux
explicitly to 2PN order ($O(v/c)^4$) and beyond.
The idea is to solve
the vacuum Einstein equations 
in the radiation zone
in an expansion 
in powers of Newton's constant $G$, and to express the
asymptotic solutions in terms of a set of formal, time-dependent,
symmetric and trace-free (STF) multipole moments \cite{thorne80}.
Then, in a near
zone within one characteristic wavelength of the radiation, the
equations including the material
source are solved in a slow-motion approximation (expansion in powers
of $1/c$) 
that yields both equations of motion for the source bodies, as well
as a set of STF source multipole moments expressed as
integrals over
the ``effective'' source, including both matter and
gravitational field contributions.
The solutions involving the
two sets of moments are then matched in an intermediate overlap zone,
resulting
in a connection between the formal radiative moments and the source
moments.
The matching also provides a natural way, using analytic continuation,
to regularize integrals involving the non-compact contributions of
gravitational stress-energy, that might otherwise be divergent.

The second group of Will, Wiseman and Pati use the approach described
in the present paper, Direct Integration of the Relaxed Einstein
Equations (DIRE), which builds upon earlier work by Epstein, Wagoner,
Will and Wiseman \cite{ew,wagwill,magnum,christo,agwtail,opus}.  
Like the PM approach,
it involves rewriting the Einstein
equations in their ``relaxed'' form, namely as an inhomogeneous,
flat-spacetime wave equation for a field $h^{\alpha\beta}$, whose
source consists of both the material stress-energy, and a
``gravitational stress-energy'' made up of all the terms non-linear in
$h^{\alpha\beta}$.  The wave equation is accompanied by a harmonic or
deDonder gauge condition
on $h^{\alpha\beta}$, which serves to specify a coordinate system, and
also imposes equations of motion on the sources.  Unlike the 
post-Minkowskian
approach, a {\it single} formal solution is written down, valid
everywhere
in spacetime.  This formal solution,
based on the flat-spacetime retarded Green function, is a retarded
integral
equation for $h^{\alpha\beta}$, which is then iterated in a
slow-motion ($v/c<1$), weak-field ($||h^{\alpha\beta}|| <1$ )
approximation, that is very similar to the corresponding procedure in
electromagnetism.  However, because the integrand of this retarded
integral
is not compact by virtue of the non-linear field contributions, one
quickly runs up against integrals that are not
well defined, or worse, are divergent.  Although at the lowest
quadrupole and first PN order, various arguments were given to
justify sweeping such problems under the rug \cite{ew,wagwill}, they 
were not very
rigorous, and provided no guarantee that the divergences would not become
insurmountable at higher PN orders.  Indeed it is straightforward to
demonstrate that at second post-Newtonian (2PN) order, the rug is
indeed pulled out from under such arguments.  

DIRE resolves these problems.  The solution of the relaxed Einstein
equation is a retarded
integral,
over the past null cone of the
field
point.  The part of the integral that extends over
the intersection between the past null cone and
the material source and the near zone
is approximated by a slow-motion expansion
involving spatial integrals of moments of the source, including the
non-compact gravitational contributions, just as in the 
post-Minkowskian and Epstein-Wagoner frameworks.
But instead of extending the
spatial integrals to infinity as was implicit in earlier
procedures, we terminate the integrals at
the boundary of the near zone, chosen to be at a radius $\cal R$ given
roughly by one wavelength of the gravitational radiation.
For the
integral over the rest of the past null cone
exterior to the near zone (``radiation zone''), 
we use a change of integration variables
to convert
the integral into a convenient, easy-to-calculate form, that is
manifestly convergent, subject only to reasonable assumptions about
the past behavior of the source, that fully accounts for the
retardation of the fields comprising the source stress-energy, and
that does not involve an explicit slow-motion expansion.
This transformation was
suggested by our earlier work on a non-linear gravitational-wave
phenomenon called the Christodoulou memory \cite{christo} (it is also
implicit in Appendix D of \cite{bd86}).
Not only are all integrations now
explicitly finite and convergent, we can show explicitly that all
contributions from the near-zone spatial integrals that depend upon
the radius
$\cal R$ are actually {\it cancelled} by corresponding terms from the
radiation-zone integrals, for all powers of $\cal R$ (including $\ln
{\cal R}$), and for any order in the PN expansion.  
Thus the procedure, as expected, has no
dependence on the arbitrarily chosen boundary radius $\cal R$
of the near-zone, and provides a simple practical method for
regularizing integrals over non-compact sources.  

The ultimate products of this work will consist of equations of
motion, gravitational waveforms, and energy flux expressions, in reasonably
ready-to-use forms.  The equations of motion for a binary system
will have the schematic
form 
\begin{equation}
d^2 {\bf x}/dt^2 = -(Gm{\bf x}/r^3)
[1+O(\epsilon)+O({\epsilon}^2)
+O({\epsilon}^{5/2})+O({\epsilon}^3)+O({\epsilon}^{7/2})
 + \dots ]\,,
\label{1-1}
\end{equation}
where $m$ is the total mass of the binary system,
${\bf x} ={\bf x}_1 -{\bf x}_2$ is the separation vector and $r=|{\bf
x}|$.
The expansion parameter $\epsilon$ is related to the orbital variables
by $\epsilon \sim Gm/rc^2 \sim (v/c)^2$, where $v$ is the relative velocity.
The leading term is
Newtonian gravity.
The next term $O(\epsilon)$ is the first
post-Newtonian correction, which
gives rise to
such effects as the advance of the periastron.
The terms of $O(\epsilon^2)$ and $O(\epsilon^3)$
are non-dissipative 2PN and 3PN corrections.  The $O(\epsilon^{5/2})$ and
$O(\epsilon^{7/2})$ terms are the leading 2.5PN and post-Newtonian
corrected 3.5PN gravitational radiation-reaction terms.
(We do not include in this discussion contributions from spin, whose
ordering in the PN hierarchy for compact bodies follows a special
convention.) Explicit formulae for terms through various orders have been
calculated by
various authors: non-radiative terms through 2PN order
\cite{damour300,DD81,Damour82,GK86,bfp98}, 
radiation reaction terms at 2.5PN and
3.5PN order \cite{iyerwill,iyerwill2,blanchet97},
and non-radiatve 3PN terms 
\cite{jaraschaefer98,jaranowski,blanchetfaye1,blanchetfaye2}

In order to derive equations of motion to the 3.5PN order shown, one must
derive the near-zone metric $g_{\alpha\beta}$ as a function of spacetime and a
functional of the source variables to 3.5PN order, which implies 
the following specific PN orders:
$g_{00}$ through $O(\epsilon^{9/2})$, 
$g_{0i}$ through $O(\epsilon^{4})$, 
$g_{ij}$ through $O(\epsilon^{7/2})$. 
In this paper we provide the required expressions in the form of
(a) Poisson-like integrals of source densities, $\int_{\cal M} 
f(t,{\bf x}^\prime )
|{\bf x}-{\bf x}^\prime|^p d^3x^\prime $, where 
$f(t,{\bf x}^\prime)$ could be proportional to source stress-energy
densities, and thus have compact support, or could be a function of
other potentials, and thus extend over the entire near-zone region of
integration $\cal M$; and (b) expressions involving time derivatives
of
source multipole
moments $M^{ijk \dots}$ contracted with spatial vectors $x^ix^jx^k
\dots$.  These expressions can be
simplified, iterated, and evaluated more explicitly, depending on the
application envisioned (``point'' mass binary system, spinning masses,
perfect fluid distributions, {\it etc.}).

The second product will be expressions for the gravitational waveform,
given schematically by
\begin{equation}
h^{ij} = {{G\mu} \over Rc^4}
\left\{ v^2 [ 1 + O(\epsilon^{1/2}) + O(\epsilon)
+ O(\epsilon^{3/2}) + O(\epsilon^2) 
+ O(\epsilon^{5/2}) + O(\epsilon^3) 
\dots ] \right\} _{TT} \,,
\end{equation}
where $\mu$ is the reduced mass, and
the subscript TT  denotes the ``transverse-traceless'' part.
The leading contribution $G\mu v^2/ Rc^4 \sim G {\ddot I}^{ij} /Rc^4$
is the standard quadrupole formula.  Explicit formulae for all terms
through 2.5PN order have been derived by various authors 
\cite{wagwill,magnum,opus,poissontail,bdi2pn,bdiww,blanchet96,blanchet98}.

From the waveform, one can also derive expressions for fluxes of
energy, angular momentum and linear momentum; the energy flux can be
written in the schematic form
\begin{eqnarray}
dE/dt = (dE/dt)_Q
[1+O(\epsilon)+O({\epsilon}^{3/2})+O({\epsilon}^2)
 +O({\epsilon}^{5/2})+O({\epsilon}^3)+ \dots ]\,,
\label{1-3}
\end{eqnarray}
%label{1-3}
where $(dE/dt)_Q$ denotes the lowest-order quadrupole contribution,

A third approach focusses on the limit in which one body is much less
massive than the other, and employs black-hole perturbation theory to
derive the gravitational waveform and energy flux, for particles
orbiting both rotating and non-rotating holes.  This method yields
both numerically accurate results as well as analytic PN expansions
up to orders as high as $(v/c)^{11}$
\cite{poissontail,ps95,cfps93,sasaki94,tagoshi94,msstt97}.  Work is currently
in progress to extend these methods beyond the test-mass
approximation, in an effort to compute corrections to first order
in $\mu/M$, the ratio of the mass of the particle to that of the
black hole \cite{wald,mino,wiseman98}.

This is the first in a series of papers that will treat the problem of
motion and gravitational radiation systematically using the DIRE
approach.  This paper lays out the foundations of the method, and
derives formal solutions to the near-zone fields through 3.5PN order
(order $(v/c)^7$ beyond Newtonian gravity), in a form useful for
future applications.  Subsequent papers in the  series will derive the
explicit equations of motion and near-zone gravitational fields for binary
systems of compact object through 2PN order, and deal with radiation reaction 
at 2.5PN and 3.5PN order.

Our conventions and notation generally follow those of
\cite{MTW,thorne80}.  Henceforth
we use units in which $G = c = 1 $.
Greek indices run over four spacetime values 0, 1, 2, 3, while
Latin indices run over three spatial values 1, 2, 3;
commas denote partial derivatives with
respect to a chosen coordinate system, while semicolons denote
covariant derivatives;
repeated indices are summed over;
$\eta^{ \mu \nu } = \eta_{ \mu \nu } = {\rm diag}(-1,1,1,1)$;
$g \equiv \det(  g_{ \mu \nu } )$;
$a^{(ij)} \equiv ( a^{ij} + a^{ji} )/2$;
$a^{[ij]} \equiv ( a^{ij} - a^{ji} )/2$;
$\epsilon^{ijk}$ is the totally antisymmetric Levi-Civita symbol
$( \epsilon^{123} = + 1)$.  We use a multi-index notation for products
of vector components and partial derivatives, and for multiple spatial
indices: 
$x^{ij \dots k} \equiv x^ix^j \dots x^k$, 
$\partial_{ij \dots k} \equiv \partial_i\partial_j \dots \partial_k$, with a
capital letter superscript denoting an abstract product of that dimensionality:
$x^Q \equiv x^{ i_1 } x^{ i_2 } ... x^{ i_q } $ and 
$\partial_Q \equiv \partial_{ i_1 } \partial_{ i_2 } ... \partial_{ i_q } $. 
Also,  for a tensor of rank $Q$, $f^Q \equiv f^{i_1 i_2 \dots i_q}$.
Angular brackets
around indices denote symmetric, trace-free (STF) combinations 
(see Appendix \ref{STF} for definitions).
Spatial indices are freely raised and lowered with
$\delta^{ij}$ and $\delta_{ij}$.

\section{Foundations of DIRE}
\label{sec:foundations}
\subsection{The relaxed Einstein equations}

We begin by reviewing the method for recasting the  
Einstein Equations
\begin{eqnarray}
R^{\alpha \beta} - {1 \over 2} g^{\alpha \beta} R = 8 \pi T^{\alpha \beta} \; ,
\label{einstein}
\end{eqnarray}
%label{einstein}
into their ``relaxed'' form.
Here $R^{\alpha \beta}$ and $R$ are the Ricci tensor and scalar,
respectively,
$g^{\alpha \beta}$ is the spacetime metric and $T^{\alpha \beta}$ 
is the stress-energy tensor of the matter.
We define the potential
\begin{eqnarray}
h^{\alpha \beta} \equiv \eta^{\alpha \beta} - (-g)^{1/2} g^{\alpha \beta} \; ,
\label{hdefinition}
\end{eqnarray}
%label{hdefinition}
(see {\it e.g.} \cite{thorne80})
and choose a particular coordinate system defined by the deDonder
or harmonic gauge condition
\begin{eqnarray}
h^{\alpha \beta},_{\beta} = 0 \; .
\label{harmonic}
\end{eqnarray}
%label{harmonic}
With these definitions 
the Einstein equations (\ref{einstein}) take the 
form
\begin{eqnarray}
\Box h^{ \alpha \beta } = -16 \pi {\tau}^{ \alpha \beta } \; ,
\label{relaxed}
\end{eqnarray}
%label{relaxed}
where $\Box \equiv  -{\partial}^2 / \partial t^2 + {\nabla}^2 $
is the flat-spacetime wave operator.
The source on the right-hand side is given by the ``effective''
stress-energy pseudotensor
\begin{eqnarray}
\tau^{\alpha\beta} = (-g)T^{\alpha\beta} + (16\pi)^{-1}
\Lambda^{\alpha\beta} \;,
\label{effective}
\end{eqnarray}
%label{effective}
where $\Lambda^{\alpha\beta}$ is the non-linear ``field'' contribution
given by
\begin{equation}
\Lambda^{\alpha \beta}
   = 16\pi (-g) t_{LL}^{\alpha \beta } 
   +  ( h^{\alpha \mu},_{\nu} h^{\beta \nu},_{\mu}
                  - h^{\alpha \beta},_{\mu \nu} h^{\mu \nu} ) \; ,
\label{nonlinear}
\end{equation}
%label{nonlinear}
and $ t_{LL}^{\alpha \beta }$ is the ``Landau-Lifshitz''
pseudotensor, given by
\begin{eqnarray}
 16 \pi (-g)t_{LL}^{\alpha \beta } &\equiv&  
 g_{\lambda\mu}g^{\nu\rho}{h^{\alpha\lambda}}_{,\nu}{h^{\beta\mu}}_{,\rho} 
+{1 \over 2} 
 g_{\lambda\mu}g^{\alpha\beta}{h^{\lambda\nu}}_{,\rho}{h^{\rho\mu}}_{,\nu} 
- 2g_{\mu\nu}g^{\lambda (\alpha}{h^{\beta )\nu}}_{,\rho}{h^{\rho\mu}}
_{,\lambda} 
\nonumber \\
&&+ {1 \over 8}
(2g^{\alpha\lambda}g^{\beta\mu}-g^{\alpha\beta}g^{\lambda\mu})
(2g_{\nu\rho}g_{\sigma\tau}-g_{\rho\sigma}g_{\nu\tau})
{h^{\nu\tau}}_{,\lambda}{h^{\rho\sigma}}_{,\mu}  \;.
\label{landau}
\end{eqnarray}
%label{landau}
By virtue of the gauge condition
(\ref{harmonic}), this source term
satisfies the conservation law
\begin{equation}
{{\tau}^{\alpha \beta}}_{, \beta} = 0 \; ,
\label{conservation}
\end{equation}
%label{conservation}
which is equivalent to the equation of motion of the matter
\begin{equation}
{T^{\alpha\beta}}_{;\beta}=0.  
\label{covariantmotion}
\end{equation}
%label{covariantmotion}

Equation (\ref{relaxed}) 
is exact, and relies only on the assumption that spacetime can be
covered by harmonic coordinates.  It is called ``relaxed'' because it
can be solved formally as a functional of source variables without
specifying the motion of the source.  Then, the harmonic gauge
condition, Eq. (\ref{harmonic}) or the equations of motion, Eq.
(\ref{covariantmotion}) are imposed to determine the metric as a
function of spacetime.

Notice that the ``source'' in Eq. (\ref{relaxed}) contains a
gravitational part that depends explicitly on
$h^{\alpha \beta}$, the very quantity for which we are trying to solve.
Also,
we can expect $\tau^{\alpha \beta}$,
which depends on the fields $h^{\alpha \beta}$, to have infinite
spatial extent.  Indeed the very outgoing radiation that we
hope to calculate, will, at some level of approximation, serve as
a contribution to the source, thus generating
an additional component of the radiation.

Another complication in Eq. (\ref{relaxed}) is that
the second derivative term $h^{\alpha \beta},_{\mu \nu} h^{\mu \nu}$
in the source really ``belongs'' on the left-hand side with the other
second derivative terms in the wave operator.
This term modifies the propagation
characteristics of the field from the flat-spacetime characteristics
represented by the d'Alembertian operator to those of the true null
cones of the curved spacetime around the source, which deviate from
the flat null cones of the harmonic coordinates.
Nevertheless, 
the DIRE technique automatically recovers the leading
manifestations of this effect, commonly known as ``tails''.

The material will be modeled as perfect fluid, having
stress-energy tensor
\begin{eqnarray}
T^{\alpha\beta} \equiv (\rho +p)u^\alpha u^\beta +pg^{\alpha\beta}
\;,
\label{fluid}
\end{eqnarray}
%label{fluid}
where $\rho$ and $p$ are the locally measured energy density and
pressure, respectively, and $u^\alpha$ is the four-velocity of an
element of fluid.  Until we begin to apply our results to specific
physical situations, such as compact binary stars, we will have no
need to specialize $T^{\alpha\beta}$ further.  

\subsection{Near-zone and radiation-zone}

We consider the material source to consist of a bound system of characteristic
size ${\cal S}$, with a suitably defined center of mass 
chosen to be at the origin of
coordinates, ${\bf X}=0$.  The {\it source zone} then consists of
the world tube ${\cal T} =\{ x^\alpha | r<{\cal S}, -\infty <t<\infty \}$.
Outside ${\cal T}$, $T^{\alpha\beta}=0$.

The fluid is assumed to move with characteristic velocity $v \ll 1$.  The
characteristic reduced wavelength  of gravitational radiation,
${\lambda\!\!\!{\scriptscriptstyle{{}^{-}}}} = \lambda /2\pi \sim
{\cal S} /v \equiv {\cal R}$ 
serves to define the boundary of the {\it near zone},
defined to be the world tube  ${\cal D} =\{x^\alpha | r<{\cal R}, -\infty
<t<\infty$ \}.  Within the near zone, the gravitational fields can be
treated as almost instantaneous functions of the source variables, {\it
i.e.}
retardation can be ignored or treated as a small perturbation of
instantaneous solutions.  For physical situations of interest, 
up to the point where 
the
post-Newtonian approximation breaks down, ${\cal R} \gg {\cal S}$.
The region exterior to the near zone is the {\it radiation zone}, $r >
{\cal R}$.  

\begin{figure}
\begin{center}
\leavevmode
\psfig{figure=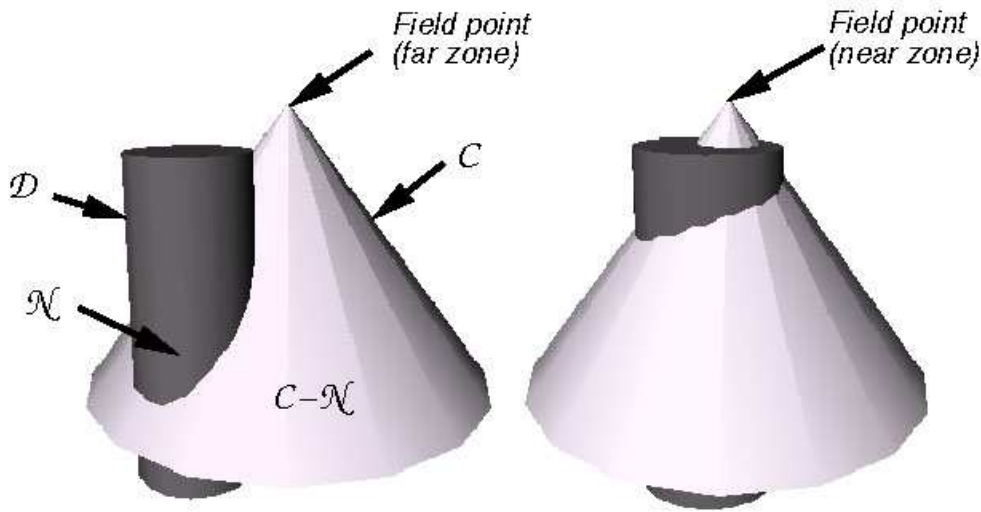,height=3.0in}
\end{center}
\caption{Past harmonic null cone $\cal C$ of the field point
intersects the near zone $\cal D$ in the hypersurface $\cal N$.  Top:
field point in the far zone; Bottom: field point in the near zone.
Inner integrals are over the hypersurface $\cal N$, and outer
integrals are over the remainder ${\cal C} - {\cal N}$ of the null
cone.
}
\label{direfig1}
\end{figure}

The formal ``solution'' to Eq. (\ref{relaxed}) with an outgoing wave boundary
condition can be written down
in terms of the retarded, flat-space Green function:
\begin{eqnarray}
h^{\alpha \beta} (t,{\bf x}) = && 4 \int
{ \tau^{\alpha \beta} (t^\prime, {\bf x^\prime} )
  \delta( t^\prime - t + | {\bf x} - {\bf x^\prime} | )
\over | {\bf x} - {\bf x^\prime} | } d^4x^\prime \;,
\label{bigintegral}
\end{eqnarray}
%label{bigintegral}
but is really just a conversion of the differential
equation (\ref{relaxed}) to an integral equation.
It represents an integration of $\tau^{\alpha \beta}/|
{\bf x} - {\bf x^\prime} |$ over the past harmonic null cone $\cal C$
emanating from the field point $(t,{\bf x})$
(see Figs. 1 and 2).  This past null cone intersects the world tube $\cal D$
enclosing
the near zone at the three-dimensional hypersurface $\cal N$.
Thus the integral of Eq. (\ref{bigintegral}) consists of two pieces,
an integration over the hypersurface $\cal N$, and an integration
over the rest of the past null cone ${\cal C} - {\cal N}$.  Each of
these integrations will be treated differently.  We will also treat
differently the two cases in which (a) the field point is
outside the near zone, and (b) the field point is within
the near zone (Fig. \ref{direfig1}).  The former case will be relevant for
calculating the gravitational-wave signal, while the latter will be
important for calculating field contributions to $\tau^{\alpha\beta}$
that must be integrated over the near zone, as well as for calculating
fields that enter the equations of motion for the source.

\subsection{Radiation-zone field point, inner integration }
\label{sec: farnear}

For a field point in the  radiation zone, and integration over the
near zone (inner integral), we 
first carry out the $t^\prime$ integration in Eq.
(\ref{bigintegral}), to obtain
\begin{eqnarray}
h_{\cal N}^{\alpha \beta} (t,{\bf x}) = && 4 \int_{\cal N} 
{ \tau^{\alpha \beta} (t -| {\bf x} - {\bf x^\prime} |, {\bf x^\prime} )
\over | {\bf x} - {\bf x^\prime} | } d^3x^\prime \;.
\label{nearintegral}
\end{eqnarray}
%label{nearintegral}
Within the near zone, the spatial integration variable ${\bf
x^\prime}$ satisfies $|{\bf x^\prime}| \le {\cal R} < r$, where the
distance to the field point $r=
|{\bf x}|$.
Expanding the $x^\prime$-dependence in both occurrences of $| {\bf x}
- {\bf x^\prime} |$ in the integrand
in powers of ${ |{\bf x^\prime }|/ r}$,  it is straightforward to show
that
\begin{eqnarray}
h_{\cal N}^{\alpha \beta} (t,{\bf x}) = &&4 \sum_{q=0}^\infty
{{(-1)^q} \over {q!}} \partial_Q \left ( {1 \over r} M^{\alpha\beta Q} 
(u) \right ) \;,
\label{genexpand}
\end{eqnarray}
%label{genexpand}
where
\begin{eqnarray}
M^{\alpha\beta Q} (u) \equiv \int_{\cal M} \tau^{\alpha\beta} (u,{\bf x^\prime})
{x^\prime}^{Q} d^3 x^\prime \;.
\label{genmoment}
\end{eqnarray}
%label{genmoment}
In Eqs. (\ref{genexpand}) and (\ref{genmoment}), the index $Q$ is a
multi-index, such that $\partial_Q \equiv
\partial_{i_1}\partial_{i_2} \dots \partial_{i_q}$ and the superscript
$Q$ in $M^{\alpha\beta Q}$ denotes $i_1 i_2 \dots i_q$, with summation
over repeated indices assumed.
The integrations in Eq. (\ref{genmoment}) are 
now over the hypersurface $\cal M$, which is the
intersection of the near-zone world-tube with the constant-time
hypersurface $t_{\cal M}=u=t-r$.
Roughly speaking, each term in the Taylor series is smaller than its
predecessor by a factor of order $v \ll 1$, provided we restrict
attention to slow-motion sources. 

Note that the field and source
variables appearing in the integrand $\tau^{\alpha\beta}$ are
evaluated at the single retarded time $u$; however,
because the field contributions to $\tau^{\alpha\beta}$ fall off as
some power of $r$, one can expect to encounter 
integrals that depend on positive powers of the radius
$\cal R$ of the boundary of integration, especially in some of the
higher-order moments.  If this boundary were to be
formally taken to $\infty$ (as has been the conventional approach in
the past), 
these integrals
would diverge.  Instead we shall demonstrate (Sec. \ref{cancel} and
Appendix \ref{cancellation}) that 
such $\cal R$-dependent
effects are {\it precisely} cancelled by contributions from the ``outer''
integral.

For the gravitational-wave signal, we need only to focus
on the spatial components of $h^{\alpha\beta}$,
and on the leading component in $1/R$, where $R$ is the distance to
the detector.  Using the fact that $u_{,i} =
-\hat N^i$,
where ${\bf \hat N} \equiv {\bf x}/R$ denotes the observation direction,
we obtain
\begin{eqnarray}
h_{\cal N}^{ij}(t, {\bf x}) = {4 \over R} \sum_{m=0}^\infty {1 \over
{m!}}
{\partial^m \over {\partial t^m}} \int_{\cal M}
\tau^{ij} (u , {\bf x^\prime} ) ({\bf \hat N \cdot x^\prime}
)^m d^3x^\prime + O(R^{-2})  \;.
\label{series}
\end{eqnarray}
%label{series}

\subsection{Radiation-zone field point, outer integration}
\label{sec: farfar}

\begin{figure}
\begin{center}
\leavevmode
\psfig{figure=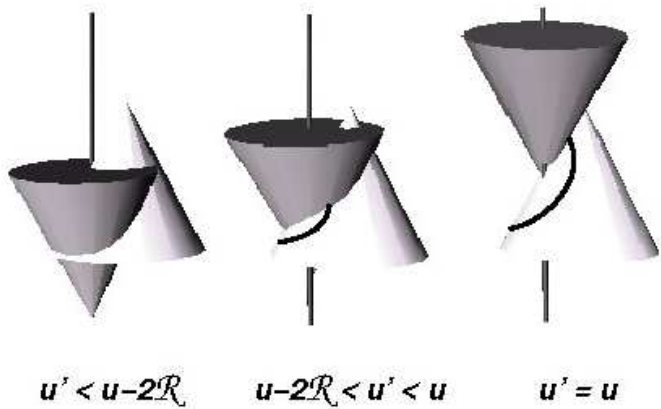,height=2.5in}
\end{center}
\caption{Change of variables for the outer integrals.  Vertical line
represents the material source world line.  The variable
$u^\prime$ is constant on the two-dimensional intersection between the
past null cone of the field point and a future null cone from the
center of mass of the system.  Left: $u^\prime < u-2{\cal R}$, 
the two cones intersect fully outside 
the near zone, so
the angular integrations are complete.  
Middle: $u-2{\cal R} < u^\prime <u$, angular integration terminates where the
intersection between the two cones meets the boundary of the near
zone.  Right: $u^\prime=u$, the upper limit of
integration; the two cones are tangent to one another.  
}
\label{direfig2}
\end{figure}

By making a change of integration variable from $(r^\prime \,,
\theta^\prime\,, \phi^\prime)$ to $(u^\prime \,,
\theta^\prime\,, \phi^\prime)$, where
\begin{equation}
t-u^\prime =r^\prime + | {\bf x} -
{\bf x^\prime} | \,,
\label{uprime}
\end{equation}
%label{uprime}
we can write
the integral over the rest of the past null cone ${\cal C} - {\cal N}$
in the form
\begin{eqnarray}
h_{{\cal C}-{\cal N}}^{\alpha\beta} (t,{\bf x}) = 
4 \int_{-\infty}^u du^\prime \oint_{{\cal C}-{\cal N}} 
{ {\tau^{\alpha\beta} (u^\prime +r^\prime, {\bf x^\prime} )} 
\over {t-u^\prime- {\bf \hat n^\prime \cdot x} } } 
[r^\prime (u^\prime, \Omega^\prime)]^2  d^2 \Omega^\prime  \;,
\label{hrest}
\end{eqnarray}
%label{hrest}
where, from Eq. (\ref{uprime})
\begin{eqnarray}
r^\prime (u^\prime, \Omega^\prime ) = [(t-u^\prime )^2
-r^2]/[2(t-u^\prime- {\bf \hat n^\prime \cdot x})] \;.
\label{rprime}
\end{eqnarray}
%label{rprime}
This change of variables represents an
integration first over the two-dimensional
intersection of the past null cone $\cal C$ with
the future null cone $t^\prime=u^\prime +r^\prime$ emanating from the
center of mass of the system at $t_{\rm CM}=u^\prime$ (Fig. \ref{direfig2}),
followed by the  $u^\prime$-integration
over all such future-directed cones, starting from the
infinite past, and terminating in
the cone emanating from the center of mass at time $u$, which is tangent to
the past null cone of the observation point.

For explicit calculations, it is useful to choose the
field point $\bf x$ to be in the z-direction, so that ${\bf \hat
n^\prime \cdot
x}=r \cos \theta^\prime$, and to write the outer integral in the form
\begin{eqnarray}
h_{{\cal C}-{\cal N}}^{\alpha\beta} (t,{\bf x}) &=& 
4 \int_{u-2{\cal R}}^u du^\prime \int_0^{2\pi} d\phi^\prime
\int_{1-\alpha}^1
{ {\tau^{\alpha\beta} (u^\prime +r^\prime, {\bf x^\prime} )} 
\over {t-u^\prime- {\bf \hat n^\prime \cdot x} } } 
[r^\prime (u^\prime, \Omega^\prime)]^2  d \cos \theta^\prime
\nonumber \\
&&+ 4 \int_{-\infty}^{u-2{\cal R}} du^\prime \oint
{ {\tau^{\alpha\beta} (u^\prime +r^\prime, {\bf x^\prime} )} 
\over {t-u^\prime- {\bf \hat n^\prime \cdot x} } } 
[r^\prime (u^\prime, \Omega^\prime)]^2  d^2 \Omega^\prime \;,
\label{houter}
\end{eqnarray}
%label{houter}
where
\begin{eqnarray}
\alpha (u^\prime) = (u-u^\prime)(2r-2{\cal R}+u-u^\prime)/2r{\cal R}  \;.
\label{alpha}
\end{eqnarray}
%label{alpha}
The incomplete angular integration in the first integral of Eq.
(\ref{houter}) reflects the fact that
for $u \ge u^\prime \ge u-2{\cal
R}$, the two-dimensional intersections meet the boundary of the near
zone.  
For $u^\prime < u-2{\cal R}$, the angular integration covers the full
$4\pi$.  
Note that $\tau^{\alpha\beta}$ contains only field contributions evaluated in
the radiation zone; because they are themselves retarded, the ``time
dependence'' $u^\prime+r^\prime = t - | {\bf x} -
{\bf x^\prime} |$ is approximately constant over each angular
integration, since it follows the hypersurface $t- | {\bf x}| = u =$
constant,
and the dominant contribution to the fields comes from $|{\bf
x^\prime} | < {\cal R}$.  This allows a kind of slow-motion, multipole
expansion to be exploited in evaluating these integrals, despite their
range well outside the near zone.

\subsection{Near-zone field point, inner integration}
\label{sec: nearnear}

In this case,
in Eq. (\ref{bigintegral}), both $\bf x$ and ${\bf x}^\prime$ are
within the near zone, hence $|{\bf x}-{\bf x}^\prime | \le 2{\cal R}$.
Consequently, the variation in retarded time can be treated as a small
perturbation, since $\tau^{\alpha\beta}$ varies on a time scale $\sim
{\cal R}$.  We therefore expand the retardation in powers of $|{\bf
x}-{\bf x}^\prime |$, to obtain
\begin{eqnarray}
h_{\cal N}^{\alpha \beta} (t,{\bf x}) = && 4 \sum_{m=0}^\infty {1
\over {m!}} {\partial^m \over {\partial t^m}} \int_{\cal M}
\tau^{\alpha \beta} (t,{\bf x}^\prime) 
 | {\bf x} - {\bf x^\prime} |^{m-1} d^3 x^\prime \;,
\label{nearexpand}
\end{eqnarray}
%label{nearexpand}
where $\cal M$ here denotes the intersection of the hypersurface $t=$
constant with the near-zone world-tube.  This version will be used for
explicit calculations of the near-zone metric for use in the equations
of motion. However an alternative formulation will be useful for
studying the $\cal R$-dependence of the inner integrals; substituting the
general Taylor expansion $| {\bf x} - {\bf x^\prime} |^{m-1} =
\Sigma_{q=0}^\infty (-1)^q (q!)^{-1} {x_<}^Q \partial_{>Q} ({r_>}^{m-1})$, 
where $<(>)$ denotes the lesser (greater) of 
$|{\bf x}|$ and $|{\bf x}^\prime |$, we
obtain
\begin{eqnarray}
h_{{\cal N}}^{\alpha \beta} (t,{\bf x}) &=& 4 \sum_{m=0}^\infty
{{(-1)^m} \over {m!}} {\partial^m \over {\partial t^m}} \sum_{q=0}^\infty
{{(-1)^q} \over {q!}} \int_{\cal
M} \tau^{\alpha\beta} (t,{\bf x}^\prime) {x_<}^Q {\partial_>}_Q
({r_>}^{m-1}) d^3x^\prime \,.
\label{nearnewversion}
\end{eqnarray}
%label{nearnewversion}

\subsection{Near-zone field point, outer integration}
\label{sec: nearfar}

The formulae from Section \ref{sec: farfar}, such as (\ref{rprime}) and
(\ref{alpha}), carry over to this case with
the result,
\begin{eqnarray}
h_{{\cal C}-{\cal N}}^{\alpha\beta} (t,{\bf x}) &=&
4 \int_{u-2{\cal R}}^{u-2{\cal R} + 2r} du^\prime \int_0^{2\pi} d\phi^\prime
\int_{1-\alpha}^1
{ {\tau^{\alpha\beta} (u^\prime +r^\prime, {\bf x^\prime} )}
\over {t-u^\prime- {\bf \hat n^\prime \cdot x} } }
[r^\prime (u^\prime, \Omega^\prime)]^2  d \cos \theta^\prime
\nonumber \\
&&+ 4 \int_{-\infty}^{u-2{\cal R}} du^\prime \oint
{ {\tau^{\alpha\beta} (u^\prime +r^\prime, {\bf x^\prime} )}
\over {t-u^\prime- {\bf \hat n^\prime \cdot x} } }
[r^\prime (u^\prime, \Omega^\prime)]^2  d^2 \Omega^\prime \;.
\label{houter2}
\end{eqnarray}
%label{houter2}
Notice that the $u^\prime$ integration ends at $u-2{\cal R}+2r$ rather
than $u$ because that corresponds to the last future null cone that
intersects points in the far zone.

\subsection{Iteration of the relaxed Einstein equations}

\begin{figure}
\begin{center}
\leavevmode
\psfig{figure=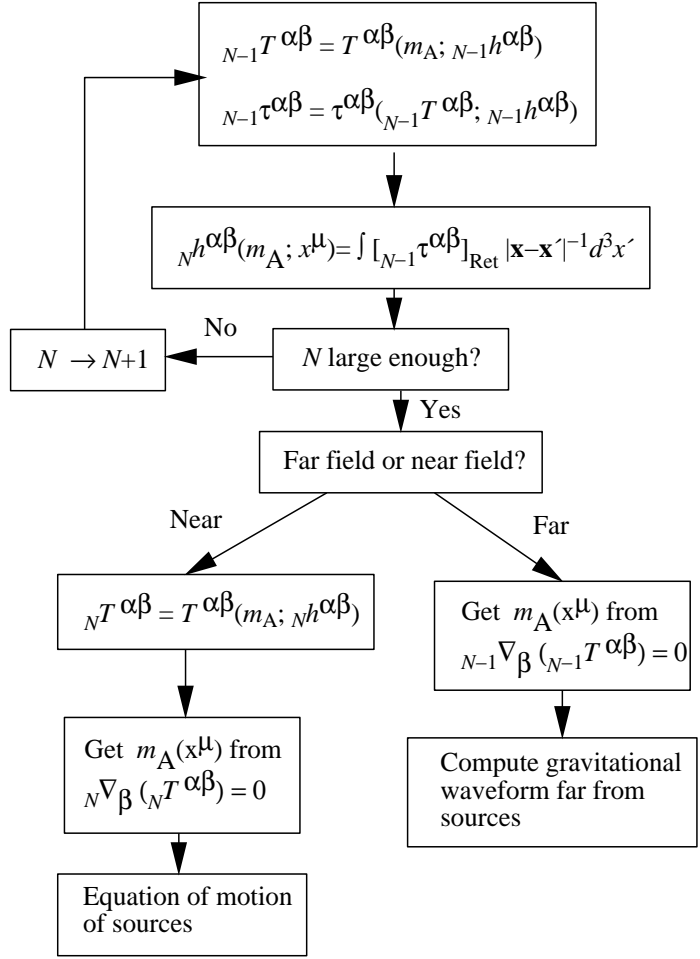,height=5.0in}
\end{center}
\caption{Structure of iteration procedure
}
\label{direfig3}
\end{figure}

Because the field $h^{\alpha\beta}$ appears in the source of the field
equation, the usual method of solution is to iterate:  substitute 
$h^{\alpha\beta}=0$ in the right-hand side of Eq. (\ref{bigintegral})
and solve for the first-iterated $_1h^{\alpha\beta}$; substitute that
into Eq. (\ref{bigintegral}) and solve for 
the second-iterated $_2h^{\alpha\beta}$, and so
on (imposing the gauge condition Eq. (\ref{harmonic}) consistently at
each order).  The general sequence of iterations is shown shematically
in Fig. \ref{direfig3}.  
The matter variables $m_A$ and the $(N-1)$-iterated field 
$_{N-1}h^{\alpha\beta}$ are used to determine $_{N-1}T^{\alpha\beta}$
and $_{N-1}\Lambda^{\alpha\beta}$.  Eq. (\ref{bigintegral}) then yields 
$_{N}h^{\alpha\beta}$ as a function of spacetime and a functional of
the matter variables.  Then, if one wishes to determine the motion of
the source, one substitutes $_{N}h^{\alpha\beta}$ into the matter
stress-energy tensor, and obtains the equations of motion from
$_N\nabla_\beta ({_NT}^{\alpha\beta}) = 0$ where $_N\nabla_\beta$ denotes
the covariant derivative using the $N$th iterated field.
If one wishes to determine the $N$th iterated 
gravitational field as a function of
spacetime (i.e. with the matter variables determined as functions of
spacetime to a consistent order), then one only needs to solve the
equations of motion $_{N-1}\nabla_\beta ({_{N-1}T}^{\alpha\beta}) = 0$,
which are equivalent to the $N$-th iterated 
gauge condition $_N{h^{\alpha\beta}}_{,\beta} = 0$.

\subsection{General structure of the outer integrals}

At the first iteration, the solution is 
simply linearized general relativity.  With $_0h^{\alpha\beta}=0$
substituted into the right-hand-side of Eq. (\ref{bigintegral}), the outer
integrals vanish, and the inner integrals over the special
relativistic $T^{\alpha\beta}$ have compact support.  There is no
$\cal R$-dependence in the integrals, trivially.  For field points
outside the source ($|\bf x| > |{\bf x}^\prime |$), within both the near and 
far zones , the first-iterated $_1h^{\alpha\beta}$ takes the form of
Eq. (\ref{genexpand}).  Since $M^{\alpha\beta Q}$ is a function only
of $u=t-r$, the spatial gradients $\partial_Q$ produce only unit
radial vectors ${\hat n}^i$, powers of $r$, and retarded time
derivatives of $M^{\alpha\beta Q}$.  Products of ${\hat n}^i$ can be
grouped into symmetric trace-free (STF) products ${\hat n}^{<L>}$, which are
analogous to $Y_{LM}$ (see Appendix \ref{STF} for useful formulae related to
STF products).  Thus, outside the source,  $_1h^{\alpha\beta}$ can be written 
as a sequence of terms of the form
\begin{equation}
{_1h^{\alpha\beta}}_{B,L}(t,{\bf x}) =  f_{B,L} (u) {\hat n}^{<L>} r^{-B}
\,.
\end{equation}

At the second iteration, in the far zone, $T^{\alpha\beta}=0$, and
$_1\Lambda^{\alpha\beta}(u^\prime + r^\prime,{\bf x}^\prime)$ 
consists of products of spatial and temporal
derivatives of $_1h^{\alpha\beta}(u^\prime + r^\prime,{\bf x}^\prime)$.  
It therefore can also be
expressed as a sequence of terms of the form 
\begin{equation}
\Lambda^{\alpha\beta}(u^\prime + r^\prime,{\bf x}^\prime) \sim
f_{B,L} (u^\prime) {\hat n}^{\prime <L>} {r^\prime}^{-B} \,.
\label{lambdadecomp}
\end{equation}
%label{lambdadecomp}
Whenever the source at a given $(N-1)$ iteration
takes this form, it is
straightforward to evaluate the general form of the outer integrals
for the $N$th iterate.
Defining the new variables $\zeta \equiv (t-u^\prime)/r =
1+(u-u^\prime)/r$, $y={\bf \hat n} \cdot {\bf \hat n}^\prime = \cos
\theta$, we
find, from Eq. (\ref{rprime}), 
\begin{equation}
r^\prime = r(\zeta^2-1)/2(\zeta -y) \,.
\label{rprime1}
\end{equation}
%label{rprime1}
Substituting Eqs. (\ref{lambdadecomp}) and (\ref{rprime1}) 
into Eq. (\ref{houter}), and
changing to integration variables $\zeta$, $y$ and
$\phi$, we obtain
\begin{equation}
{_Nh^{\alpha\beta}_{{\cal C}-{\cal N}}}_{B,L} = {1 \over 2} \left ( 
{2 \over r} \right )^{B-2} {\hat n}^{<L>} \int_{-1}^1 P_L(y)dy
\int_{\zeta(y)}^\infty {{(\zeta-y)^{B-3}} \over {(\zeta^2-1)^{B-2}}} 
f_{B,L} (u -r(\zeta-1)) d\zeta \,,
\label{outerfar0}
\end{equation}
%label{outerfar0}
where $\zeta(y)=z+\sqrt{z^2-2zy+1}$, $z={\cal R}/r$, and $P_L(y)$
is the Legendre polynomial. 

For far-zone field points, $z<1$; Taylor
expanding $f_{B,L} (u -r(\zeta-1))$ about $u$, 
we obtain, for $B>2$,
\begin{equation}
{_Nh^{\alpha\beta}_{{\cal C}-{\cal N}}}_{B,L} =  \left ({2 \over r}
\right ) ^{B-2}
{\hat n}^{<L>} \sum_{q=0}^\infty {\cal D}^q_{B,L} (z) r^q {{d^q
f_{B,L}(u)} \over {du^q}} \,,
\label{outerfar}
\end{equation}
%label{outerfar}
where the coefficients ${\cal D}^q_{B,L} (z)$
are given by 
\begin{equation}
{\cal D}^q_{B,L} (z) = {(-)^q \over q!} \int_1^{1+2z} {{(\zeta -1)^q}
\over {(\zeta^2-1)^{B-2}}} A_{B,L}(\zeta,\alpha) d\zeta
- \sum_{p=0}^q k_{B,L}^{(q-p+1)}(1+2z) {(-2z)^p \over p!} \,,
\label{calD}
\end{equation}
%label{calD}
where
\begin{mathletters}
\label{outerdefinitions}
\begin{eqnarray}
A_{B,L}(\zeta,\alpha) \equiv& {1 \over 2} \int_{1-\alpha}^1 P_L(y)
(\zeta -y)^{B-3} dy \,, \\
\alpha \equiv& (\zeta-1)(\zeta+1-2z)/2z \,, \\
dk_{B,L}^{(m)}(\zeta) / d\zeta \equiv& k_{B,L}^{(m-1)}(\zeta)  \,,\quad m \ge 1
\,; \\
\quad k_{B,L}^{(0)}(\zeta) \equiv& A_{B,L}(\zeta,2) / (\zeta^2-1)^{B-2} \,.
\end{eqnarray}
\end{mathletters}
%label{outerdefinitions}
The case $B=2$ is special,
and leads to the result
\begin{equation} 
{_Nh^{\alpha\beta}_{{\cal C}-{\cal N}}}_{2,L} = {{\hat n}^{<L>} \over
r} \int_0^\infty f_{2,L}(u-s) Q_L \left ( 1+{s \over r} \right
) ds + {\hat n}^{<L>} \sum_{q=0}^\infty {\cal D}^q_{2,L} (z)
r^q {{d^q
f_{2,L}(u)} \over {du^q}} \,,
\label{outerfar2}
\end{equation}
%label{outerfar2}
where
\begin{equation}
{\cal D}^q_{2,L} (z) = {(-)^{q+1} \over 2q!} \int_1^{1+2z} (\zeta -1)^q
d\zeta \int_{-1}^{1-\alpha} {P_L(y) \over {(\zeta -y)}} dy \,,
\label{calD2}
\end{equation}
%label{calD2}
where $Q_L(y)$ is the Legendre function.
 
Notice that, for $B \ne 2$, the outer integral returns a result of the
same generic form as the input function.  The case $B=2$ returns terms
with a logarithmic dependence on $r$ (via the $Q_L$'s); terms of this
form are called
``tails''.

Similarly, for field points in the near zone, $z>1$, we Taylor expand
$f_{B,L} (u -r(\zeta-1))$ about $u+r=t$, and obtain, for $B>2$,
\begin{equation}
{_Nh^{\alpha\beta}_{{\cal C}-{\cal N}}}_{B,L} =  \left ( {2 \over r}
\right ) ^{B-2}
{\hat n}^{<L>} \sum_{q=0}^\infty {\cal E}^q_{B,L} (z) r^q {{d^q
f_{B,L}(t)} \over {dt^q}} \,,
\label{outernear}
\end{equation}
%label{outernear}
where the coefficients ${\cal E}^q_{B,L} (z)$
are given by
\begin{equation}
{\cal E}^q_{B,L} (z) = {(-)^q \over q!} \int_{2z-1}^{2z+1} {\zeta^q
\over {(\zeta^2-1)^{B-2}}} A_{B,L}(\zeta,\alpha) d\zeta
- \sum_{p=0}^q k_{B,L}^{(q-p+1)}(1+2z) {(-1-2z)^p \over p!} \,,
\label{calE}
\end{equation}
%label{calE}
and, for $B=2$,
\begin{equation} 
{_Nh^{\alpha\beta}_{{\cal C}-{\cal N}}}_{2,L} = {{\hat n}^{<L>} \over
r} \int_0^\infty f_{2,L}(u-s) Q_L \left ( 1+{s \over r}
\right) ds + {\hat n}^{<L>} \sum_{q=0}^\infty {\cal E}^q_{2,L} (z)
r^q {{d^q
f_{2,L}(t)} \over {dt^q}} \,,
\label{outernear2}
\end{equation}
%label{outernear2}
where
\begin{equation}
{\cal E}^q_{2,L} (z) = {(-)^{q+1} \over q!} \left \{ \int_1^{2z-1} \zeta^q
Q_L(\zeta) d\zeta + {1 \over 2}
 \int_{2z-1}^{2z+1} \zeta^q d\zeta
\int_{-1}^{1-\alpha} {P_L(y) \over {(\zeta -y)}} dy \right \}\,.
\label{calE2}
\end{equation}
%label{calE2}
Notice that, for near-zone field points, the functions $f_{B,L}$ are
evaluated at the local time $t$, not retarded time $u$.

\subsection{Cancellation of $\cal R$ dependence}
\label{cancel}

It is evident that the inner integrals and outer integrals for the
field $h^{\alpha\beta}$ will separately depend upon the radius $\cal
R$ of the boundary between the near zone and the far zone.  But since
each integral was simply a rewriting of a piece of the original
integral, Eq. (\ref{bigintegral}), which had no $\cal R$ dependence, it is
equally evident that the separate $\cal R$-dependences must cancel
between the inner and outer integrals.  In \cite{opus}, referred to
hereafter as WW, we
demonstrated such a cancellation explicitly for contributions to the
gravitational waveform at 2PN order that depended on positive powers
of $\cal R$.  Here we demonstrate the cancellation generally, for both
near-zone and far-zone field points, for arbitrary powers of $\cal R$
(including $\ln {\cal R}$) and to an order of
iteration sufficient for our purposes.

The proof proceeds by induction.  First, as we pointed out above, the
first-iterated
field $_1 h^{\alpha\beta}$ is trivially independent of $\cal R$.

Secondly, we assume that the $(N-1)$-iterated field does not depend on
$\cal R$, i.e. that all $\cal R$-dependence cancels at this order of
iteration.  We wish to demonstrate that this implies cancellation of
$\cal R$-dependence in the $N$-iterated field.  The proof 
consists of considering the limiting behavior of the inner
and outer integrals for $_N h^{\alpha\beta}$ 
in the vicinity of $|{\bf x}^\prime |\to {\cal
R}$.  Here $T^{\alpha\beta}$ vanishes, and we only need to consider
$_{N-1} \Lambda^{\alpha\beta}$, which is a functional of 
$_{N-1} h^{\alpha\beta}$.
We have already seen that, in the far zone,
$_{N-1} \Lambda^{\alpha\beta}$ can be decomposed  into terms of the form
$f_{B,L}(u) {\hat n}^{<L>} r^{-B}$.
(We consider tail contributions with $\ln r$ dependence separately.)  
Since the $N-1$ iterated field does not depend on $\cal R$ by
assumption, continuity of the fields means that $_{N-1}
\Lambda^{\alpha\beta}$ will have this same form just inside the near
zone.  Thus we will calculate the limiting behavior of the inner
integral of a term of this form as the integration variable $r^\prime
\to {\cal R}$ from below, 
and compare its ${\cal R}$ dependence with that of the
outer integral of the same term.

For far-zone field points, we must calculate the ${\cal R}$ dependence
of the moments $M^{\alpha\beta Q}$, and substitute into Eq. (\ref{genexpand});
after considerable algebra (see Appendix \ref{cancellation}), 
we obtain, for the limiting
behavior of $_{N} h_{\cal N}^{\alpha\beta}$ as the integration
variable approaches $\cal R$ from inside, 
\begin{equation}
{_Nh^{\alpha\beta}_{\cal N}}_{B,L} 
\to \left ({2 \over r}
\right ) ^{B-2}
{\hat n}^{<L>} \sum_{q=0}^\infty {\cal D}^{{\rm in},q}_{B,L} (z) r^q {{d^q
f_{B,L}(u)} \over {du^q}} \,,
\label{farinnerlimit}
\end{equation}
%label{farinnerlimit}
where
\begin{eqnarray}
{\cal D}^{{\rm in},q}_{B,L} (z) &=& \sum_{m=0}^q \sum_{j=0}^{j_{max}}
 {{(-)^m 2^{2+j-B}} \over {m!(q-m-j+2L+1)!}} {{[{1 \over
2}(q-m-j+2L)]!} \over {[{1 \over 2}(q-m-j)]!}}
{{(2L-j)!} \over {j!(L-j)!}} \nonumber \\
&& \qquad \times \left \{ \begin{array}{ll}
{z^{3+L-B+q-j} / (3+L-B+q-j)} &  \quad 3+L-B+q-j \ne 0 \\ 
\ln {\cal R}  &\quad  3+L-B+q-j =0 \,. \end{array} 
\right .
\label{farinnerlimit2}
\end{eqnarray}
%label{farinnerlimit2}
where $j_{max} = \, {\rm lesser \, of} \, \{q-m,L\}$, and $q-m-j = \, {\rm
even \, integer} \, \ge 0$.
Eq. (\ref{farinnerlimit}) is of the same form as the 
outer integral for far-zone field
points, Eq. (\ref{outerfar}).  The coefficients ${\cal D}^{q}_{B,L}
(z)$ from the outer integrals are most easily evaluated using computer
algebra methods (we calculated the coefficients using independent
Maple and Mathematica programs); the result is, for each $B$, $L$ and $q$, 
\begin{equation}
{\cal D}^{{\rm in},q}_{B,L} (z) + \{ z \,{\rm -dependent \, part \, of}\,
{\cal D}^{q}_{B,L} (z) \} = 0\,.
\end{equation}
Thus the $\cal R$ dependence cancels term by term.  Similar
cancellation occurs for near-zone field points, as well as for the
case where the integrand has $r^{-B} \ln r$ dependence.  Details are
given in Appendix \ref{cancellation}.

This cancellation, while inevitable, has practical consequences, in
the following sense.  In calculating the inner contributions to the 
fields, we must integrate
over a finite hypersurface, $\cal M$, sources that extend throughout
$\cal M$.  Consequently, any such integral will have terms that are
independent of $\cal R$, as well as terms that depend
on ${\cal R}^q$ or $\ln {\cal R}$.  Because we know that all terms of
the latter form cancel
with contributions from the outer integrals 
in the final expression for the field, we can drop them in any
individual result.
Similarly, we can drop all $\cal R$-dependent terms that arise in any
individual outer integral.  This provides a kind of regularization of
integrals, that cures the problem of divergent integrals that haunted earlier
slow-motion methods.  In fact, one can show that there is a close
connection between this method of regularization and the method of
analytic continuation used by Blanchet  \cite{Lucproof}.

Thus, our procedure for determining the field is to determine
separately $h^{\alpha\beta}_{\cal N}$ and 
$h^{\alpha\beta}_{{\cal C}-{\cal N}}$
to a given PN order,
keeping only $\cal R$-independent terms in each expression, then sum
them to obtain
\begin{equation}
h^{\alpha\beta} = h^{\alpha\beta}_{\cal N}+h^{\alpha\beta}_{{\cal
C}-{\cal N}}\,.
\end{equation}

\section{Weak field, slow-motion approximation}

We now turn to a discussion of the number $N$ of iterations 
needed to derive equations
of motion or gravitational waveforms of a desired accuracy, in a
weak-field, slow-motion approximation.

We assume that, for the fluid source, 
\begin{eqnarray}
v^2 \sim m/{\cal S} \sim p/\rho \sim \epsilon \ll 1 \;,
\end{eqnarray}
where $\epsilon$ will be used as an expansion parameter.  But from the
nature of the iteration procedure, it is evident that each iteration
of the field introduces corrections of order $m/{\cal S}$.  In terms of
$\epsilon$, $m$ and $\cal S$
the equations of motion (\ref{1-1}) can be rewritten schematically as
\begin{equation}
d {\bf v}/dt \sim (m/{\cal S}^2)
[1+O(\epsilon)+O({\epsilon}^2)+O({\epsilon}^{5/2})+O({\epsilon}^3)
+O({\epsilon}^{7/2}) + \dots ]\,,
\end{equation}
where the terms inside the square brackets represent the Newtonian,
post-Newtonian, 2PN, 2.5PN (radiation-reaction), 3PN, and 3.5PN
(radiation-reaction)
terms respectively. 
For a term of order $\epsilon^N$, the largest
number of powers of $m/{\cal S}$ that can appear in it (including one power
from the $m/{\cal S}^2$ prefactor) is $N+1$.  The radiation reaction terms of order
$\epsilon^{N+1/2}$ must contain
an odd number of velocities (in order to be odd under time-reversal), 
thus the maximum number of powers of
$m/{\cal S}$ for them is also $N+1$.  Since one iteration gives the
Newtonian potential, which yields the Newtonian equations of motion
($N=1$), then, to obtain the 1PN terms ($N=2$), one must
have the second iterated field, to obtain the 2PN and 2.5PN terms  
($N=3$), one must have the
third iterated field, while to obtain the 3PN and 3.5PN terms ($N=4$),
one must have the fourth iterated field. 

Similarly,
to obtain a result for the waveform accurate to the
order of the quadrupole formula, $h \sim \ddot {\cal I}^{ij}/R \sim (m/R)
(v^2 + m/{\cal S}) \sim \epsilon^2$ ($N=2$), the second-iterated field
is needed.  Note that the term $m/{\cal S}$ in $\ddot {\cal I}^{ij}$
arises through the use
of the Newtonian equation of motion.
Then, to obtain the 1PN, 2PN and 3PN
corrections to the quadrupole approximation, 
the third, fourth, and fifth-iterated fields are needed,
respectively.  This
would be an impossible task, if it weren't for the judicious
use of the conservation
law,
Eq. (\ref{conservation}).  Consider for example, 
the source $_{N-1} \tau^{ij}$ of the $N$th iterated
gravitational-wave field $_Nh_{\cal N}^{ij}$, Eq. (\ref{series}),
specifically the leading, $m=0$ term.  The conservation
law, Eq. (\ref{conservation}), 
converts $_{N-1} \tau^{ij}$ into two time derivatives of
$_{N-1} \tau^{00}x^ix^j$ (modulo total divergences).  Because of the 
slow-motion approximation, two time derivatives increase the order by
$\epsilon$, and thus, to sufficient accuracy, only 
$_{N-2} \tau^{00}$ is needed in practice.  An important caveat to this
is that the surface terms
that arise from the total divergences and the outer integrals
must formally be evaluated using the
$N-1$ expressions.  However, in practice, these terms contribute at
sufficiently high order that 
they can be treated without resort to explicit
$N-1$ expressions.   Effectively, the burden of
accuracy has been shifted from the $N$th-iteration of the field, to the
$N-1$-iterated equations of motion, which enter via the two
time-derivatives, and which are needed anyway to evaluate the field as a
function of spacetime.  Thus, for $N=2$, the leading quadrupole
approximation, only $_0 \tau^{00} = \rho$ is needed, together with the
Newtonian equations of motion.
This circumstance is responsible for the
prevalent, but erroneous view that linearized gravity (one iteration)
suffices to derive the quadrupole formula.  The formula so derived 
turns out to
be ``correct'', but its foundation is not (see \cite{walkerwill} for
discussion).

Thus, in WW, to evaluate the 2PN waveforms (fourth iteration),
only second-iterated fields were
needed in the source terms.
For 3PN waveforms, only third-iterated
fields will be needed.

\section{Formal structure of Near-Zone Fields}

\subsection{Metric and stress-energy pseudotensor in terms of the
fields}

We begin by defining a simplified notation for the field
$h^{\alpha\beta}$:
\begin{eqnarray}
N &\equiv& h^{00} \sim O(\epsilon) \,, \nonumber \\
K^i &\equiv& h^{0i} \sim O(\epsilon^{3/2}) \,, \nonumber \\
B^{ij} &\equiv& h^{ij} \sim O(\epsilon^2) \,, \nonumber \\
B &\equiv& h^{ii} \equiv \sum_i h^{ii} \sim O(\epsilon^2) \,, 
\end{eqnarray}
where we show the leading order dependence on $\epsilon$ in the near
zone.  To obtain
the equations of motion to 3.5PN order, we need to determine
the components of the physical metric to the following orders: 
$g_{00}$ to $O(\epsilon^{9/2})$, 
$g_{0i}$ to $O(\epsilon^{4})$ , and
$g_{ij}$ to $O(\epsilon^{7/2})$.
From the definition (\ref{hdefinition}), one can invert to find
$g_{\alpha\beta}$ in terms of $h^{\alpha\beta}$.  Expanding to the
required order, we find,
\begin{mathletters}
\label{metricexpand}
\begin{eqnarray}
g_{00} &=& -(1- {1 \over 2}N+ {3 \over 8} N^2 - {5 \over 16} N^3 + {35
\over 128}N^4 ) + {1 \over 2}B(1- {1 \over 2}N+ {3 \over 8} N^2)
\nonumber \\
&&+ {1 \over 4}(B^{ij}B^{ij}-{1 \over 2}B^2) 
+{1 \over 2}K^jK^j -  {3 \over 4} NK^jK^j +O(\epsilon^5) \,, \\
g_{0i} &=& -K^i(1- {1 \over 2}N -{1 \over 2}B + {3 \over 8} N^2) - K^j
B^{ij} +O(\epsilon^{9/2}) \,, \\
g_{ij} &=& \delta^{ij} (1+ {1 \over 2}N- {1 \over 8} N^2 + {1 \over 16}
N^3 - {1 \over 4}NB +{1 \over 2}K^kK^k ) \nonumber \\
&&+ B^{ij} - {1 \over
2}B\delta^{ij}- K^iK^j + {1 \over 2}N B^{ij} +O(\epsilon^4) \,, \\
(-g) &=& 1+N-B-NB+K^iK^i + O(\epsilon^4) \,.
\end{eqnarray}
\end{mathletters}
%label{metricexpand}
Notice that, in order to find the metric $g_{\alpha\beta}$ to the
desired order, we must obtain $N$ to $O(\epsilon^{9/2})$, $K^i$ to
$O(\epsilon^{4})$, $B^{ij}$ to $O(\epsilon^{7/2})$ and $B$ to
$O(\epsilon^{9/2})$.  In fact, because $B$ contributes linearly to
$g_{00}$, we will treat $B$ differently from $B^{ij}$.

Using Eq. (\ref{metricexpand}), we can express the
matter stress-energy  tensor $T^{\alpha\beta}$, Eq. (\ref{fluid}),
as a PN
expansion.  However, the details of such an expansion
will depend on the basic variables used to characterize
the matter.  For example, to discuss the structure of a
star in a PN expansion, it is convenient to use the
mass-energy density $\rho$ and pressure $p$, together
with an equation of state.  However, to discuss the
motion of compact bodies in an effective ``point-mass''
limit, it is more convenient to ignore the pressure
totally, and to use the so-called ``conserved'', or
baryon density, $\rho^* \equiv \rho \sqrt{-g} u^0$.
For now, we follow the convention of Damour {\it et
al.}, and define the quantities
\begin{eqnarray}
\sigma & \equiv& T^{00} + T^{ii} \,, \nonumber \\
\sigma^i & \equiv& T^{0i} \,, \nonumber \\
\sigma^{ij} & \equiv& T^{ij} \,.
\end{eqnarray}
We will express various potentials 
formally in terms of these
densities, and later make a PN expansion of them in
terms of the densities most appropriate to the
application.

Substituting the formulae for $h^{\alpha\beta}$ and
$g_{\alpha\beta}$ into Eqs. (\ref{nonlinear}) and (\ref{landau}) for
$\Lambda^{\alpha\beta}$, we obtain, to the required order,
\begin{mathletters}
\label{Lambda}
\begin{eqnarray}
\Lambda^{00} &=& -{7 \over 8} (\nabla N)^2 \nonumber \\
&&+ \left \{ {5 \over 8} {\dot N}^2 - \ddot N N -2 {\dot N}^{,k} K^k
+ {1 \over 2} K^{i,j} (3 K^{j,i}+ K^{i,j}) \right . \nonumber \\
&& \left . + {\dot K}^j N^{,j} - B^{ij} N^{,ij} + 
{1 \over 4} \nabla N \cdot
\nabla B + {7 \over 8} N (\nabla N)^2 \right \} \nonumber \\
&&+ \left \{ K^{k,j} {\dot B}^{jk} +{1 \over 4} B^{jk,l}
(B^{jk,l} -2B^{kl,j}) + {1 \over 4} \dot N \dot B - {1
\over 8} (\nabla B)^2  +{1 \over 4} \dot N N^{,j}K^j \right . \nonumber \\
&&  \left . 
+{7 \over 8} N^{,j}N^{,k} B^{jk}
-{1 \over 2} K^j N^{,k} (3 K^{j,k}+ 4K^{k,j}) -{7 \over
8} N^2 (\nabla N)^2 \right \} + O(\rho \epsilon^4) \,,
\\
\Lambda^{0i} &=& \left \{ N^{,k}( K^{k,i}- K^{i,k}) +{3 \over 4}\dot
N N^{,i} \right \} \nonumber \\
&&+ \left \{ \dot N {\dot K}^i - N {\ddot K}^i -2 K^k
{\dot K}^{i,k} - B^{lm} K^{i,lm} + K^{k,l} (B^{il,k}
+B^{ik,l} -B^{kl,i})+ N^{,k} {\dot B}^{ik} \right . \nonumber \\
&& \left .  - {1 \over 4} \dot N B^{,i} 
-{1 \over 4} N^{,i} \dot B - NN^{,k}( K^{k,i}- K^{i,k}) -
{3 \over 4} N \dot N N^{,i} + {1 \over 8} K^i (\nabla
N)^2 -{1 \over 4}K^k N^{,k} N^{,i} \right \} \nonumber
\\
&&+ O(\rho
\epsilon^{7/2}) \,, \\
\Lambda^{ij} &=& {1 \over 4} \{ N^{,i}N^{,j} - {1 \over 2}
\delta^{ij} (\nabla N)^2 \} \nonumber \\
&&+ \left \{ 2 K^{k,(i}K^{j),k}- K^{k,i}K^{k,j}
-K^{i,k}K^{j,k} +2N^{,(i} {\dot K}^{j)} +{1 \over
2}N^{,(i} B^{,j)} \right . \nonumber \\
&& \left . -{1 \over 2}N(N^{,i}N^{,j} - {1 \over 2}
\delta^{ij} (\nabla N)^2) -\delta^{ij} (K^{l,k}K^{[k,l]} 
+N^{,k}{\dot K}^{k}
+{3 \over 8}{\dot N}^2 +{1 \over 4} \nabla N \cdot
\nabla B ) \right \} \nonumber \\
&&+ \left \{ 2 {\dot K}^i{\dot K}^j + {\dot
B}^{k(i}(K^{j),k}-K^{k,j)})-2{\dot B}^{ij,k} K^k
-N{\ddot B}^{ij} -B^{ij,lm} B^{lm} \right .\nonumber \\
&& + B^{ik,l} ( B^{jl,k}+ B^{jk,l}) -2B^{kl,(i} B^{j)k,l}
+\frac{1}{2} B^{kl,i}B^{kl,j} - \frac{1}{4}B^{,i}B^{,j}
\nonumber \\
&& -N( 2 K^{k,(i}K^{j),k}- K^{k,i}K^{k,j}
-K^{i,k}K^{j,k}) +K^kK^{k,(i}N^{,j)} -2NN^{,(i}{\dot
K}^{j)} - {1 \over 2} \dot N N^{,(i} {K}^{j)} \nonumber
\\
&& -{1 \over 2}N^{,k}  N^{,(i}B^{j)k} -{1 \over
2}NN^{,(i} B^{,j)} +{1 \over 8} (\nabla N)^2 B^{ij}
+{3 \over 4} N^2( N^{,i}N^{,j} - {1 \over 2}
\delta^{ij} (\nabla N)^2) \nonumber \\
&& +{1 \over 8}\delta^{ij} [ (\nabla B)^2 +2 \dot N
\dot B +8K^{k,l}{\dot B}^{kl}+4B^{kl,m}B^{km,l}
-2B^{kl,m}B^{kl,m}+3N{\dot N}^2 +8NN^{,k}{ \dot K}^{k}
\nonumber \\
&& \left . +8NK^{l,k}K^{[k,l]} -4K^kN^{,l}K^{k,l}
+2\dot N K^kN^{,k} + N^{,k}N^{,l}B^{kl} 
+2N\nabla N \cdot \nabla B ] \right \} \nonumber \\
&& + O(\rho \epsilon^4) \,,
\\
\Lambda^{ii} &=& -{1 \over 8} (\nabla N)^2 \nonumber
\\
&&+ \left \{ K^{l,k}K^{[k,l]}-N^{,k}{ \dot K}^{k} -{1
\over 4}\nabla N \cdot \nabla B -{9 \over 8} {\dot N}^2
+{1 \over 4}N ({\nabla N})^2 \right \} \nonumber \\
&&+ \left \{ 2 {\dot K}^k{\dot K}^k -2{\dot B}^{,k}K^k
+3{\dot B}^{kl}K^{k,l} -N \ddot B +{3 \over 4}\dot N
\dot B +{1 \over 8}({\nabla B})^2 -B^{,lm}B^{lm} \right . \nonumber \\
&& + {3 \over 4}B^{kl,m}B^{kl,m} +
{1 \over 2}B^{kl,m}B^{km,l}-NK^{l,k}K^{[k,l]}-
{1 \over 2}N^{,l}K^kK^{k,l} +NN^{,k}{ \dot K}^{k} \nonumber \\
&& \left . + 
{1 \over 4}\dot N N^{,k}K^k - {1 \over 8}N^{,k}N^{,l}B^{kl}
+{1 \over 4}N\nabla N \cdot \nabla B+{1 \over 8}
({\nabla N})^2 B +{9 \over 8}N{\dot N}^2 -{3 \over 8}
N^2 ({\nabla N})^2 \right \} \nonumber \\
&& +  O(\rho \epsilon^4) \,,
\end{eqnarray}
\end{mathletters}
%label{Lambda}
where an overdot denotes $\partial/\partial t$.  In the above
expressions, terms grouped
within braces make leading contributions of the same order.  For
example, in
$\Lambda^{00}$, the three groupings correspond to $O(\rho \epsilon)$,
$O(\rho \epsilon^2)$, and $O(\rho \epsilon^3)$, respectively.

\subsection{Source moments and other integral quantities}

Throughout our calculations, a number of integrals of
the source stress-energy pseudotensor occur, for
example, in the multipole expansions of Eq.
(\ref{genmoment}).  It is useful to define and collect
these quantities, and to discuss their properties.  All
integrals are carried out over a constant time (or
constant retarded time) hypersurface $\cal M$, within the
near-zone.  In general, these integrals will have $\cal
R$ dependence, but, in line with the foregoing
discussion, we shall consistently drop such
terms.  The relevant integrals are:
\begin{mathletters}
\begin{eqnarray}
P^\mu &\equiv&  M^{\mu 0} = \int_{\cal M} \tau^{\mu 0} d^3x \,, \\
{\cal I}^Q &\equiv& M^{00Q} = \int_{\cal M} \tau^{00} x^Q d^3x \,,
\\
{\cal J}^{iQ} &\equiv& \epsilon^{iab} M^{0baQ} = \epsilon^{iab}\int_{\cal M} \tau^{0b} 
x^{aQ} d^3x \,,
\\
{\cal P}^{ijabQ} &\equiv& \int_{\cal M} x^{[a}\tau^{i][j} 
x^{b]Q} d^3x \,.
\end{eqnarray}
\end{mathletters}
By making use of the equations of motion
${\tau^{\alpha\beta}}_{,\beta}=0$, we can transform
some of these integrals into other forms, modulo
surface integrals at the boundary $\partial {\cal M}$
of the near zone.  For example,
\begin{eqnarray}
{\dot P}^\mu &=& - \oint_{\partial {\cal M}} \tau^{\mu j} 
d^2 S_j \,, \nonumber \\
{\dot {\cal J}}^i &=&  - \epsilon^{iab} \oint_{\partial {\cal M}}
\tau^{jb} x^a d^2 S_j \,, \nonumber \\
{\dot {\cal I}}^i &=& P^i - \oint_{\partial {\cal M}}
\tau^{0j} x^i d^2 S_j \,.
\end{eqnarray}
These identities express the conservation of total
energy, momentum and angular momentum, and uniform
center-of-mass motion, modulo a flux of gravitational
radiation from the system.  In calculations, the
surface terms must be checked carefully to see if they
make ${\cal R}$-independent contributions to the order
considered.  For the most part, such surface terms turn
out to make no contribution.

Henceforth, we shall set ${\cal I}^i =
{\dot{\cal I}}^i =0$, which amounts to attaching the origin of
coordinates to the
center of mass of the system.  

Other useful identities include:
\begin{mathletters}
\begin{eqnarray}
M^{ij} &=& {1 \over 2} \ddot {\cal I}^{ij} 
+ {1 \over 2} \oint_{\partial {\cal M}}
[\tau^{lm}(x^{ij})_{,l} + {\dot \tau}^{m0} x^{ij} ]
d^2S_m \,,  \\
M^{ijk}&=& {1 \over 6} \ddot {\cal I}^{ijk} + 
{2 \over 3} \epsilon^{lk(i} {\dot {\cal J}}^{l|j)}
\nonumber \\
&& + {1 \over 6} \oint_{\partial {\cal M}}
[\tau^{lm}(x^{ijk})_{,l} + {\dot \tau}^{m0} x^{ijk} ]
d^2S_m - {2 \over 3} \oint_{\partial {\cal M}} 
[ \tau^{l[k} x^{i]j} + \tau^{l[k} x^{j]i} ] d^2S_l
\,, \\
M^{ijQ} &=& {1 \over {(q+1)(q+2)}} {\ddot {\cal
I}}^{ijQ} + {2 \over {(q+2)}} \epsilon^{mk_1(i} 
{\dot {\cal J}}^{m|j)k_2\dots k_q}{\rm (sym \,k:Q)} 
+{{8(q-1)} \over {(q+1)}} {\cal P}^{ij(k_1k_2\dots
k_q)} \nonumber \\
&& + {1 \over {(q+1)(q+2)}} \oint_{\partial {\cal M}}
[\tau^{lm}(x^{ijQ})_{,l} + {\dot \tau}^{m0} x^{ijQ} ]
d^2S_m \nonumber \\
&& - {2 \over {(q+2)}} \oint_{\partial {\cal M}}
[ \tau^{l[k_1} x^{i]jk_2\dots k_q} +  \tau^{l[k_1} x^{j]ik_2\dots k_q}
] d^2S_l
{\rm (sym \,k:Q)}
\,, \\ 
M^{0jQ} &=& {1 \over {q+1}} {\dot {\cal I}}^{jQ} - {q \over
{q+1}} \epsilon^{mj(k_1} {\cal J}^{m|k_2\dots k_q)} 
+ {1 \over {q+1}} \oint_{\partial {\cal M}} \tau^{0m} x^{jQ}
d^2S_m \,, 
\end{eqnarray}
\end{mathletters}
where the notation (sym k:Q) means symmetrize on the
indices $k_1$ through $k_q$, and the superscript notation
$^{m|k\dots)}$ means that only the indices following the vertical line
are involved in
symmetrization.

\subsection{Near-zone field expanded to 3.5 PN order}

We now carry out the explicit expansion of the near-zone
field through 3.5PN order, beginning with the 
inner integral, Eq. (\ref{nearexpand}), 
applying the above identities where possible.  Inserting powers
of $\epsilon$ to indicate the leading order of each term, we obtain
the result 
\begin{mathletters}
\label{bigexpansion}
\begin{eqnarray}
% 
%HERE IS N
% 
N_{\cal N} &=& 4 \epsilon \int_{\cal M} {\tau^{00}(t,{\bf x}^\prime) \over {|{\bf x}-{\bf
x}^\prime |}} d^3x^\prime 
+2 \epsilon^2 \partial^2_t \int_{\cal M} \tau^{00}(t,{\bf x}^\prime) |{\bf x}-{\bf x}^\prime | d^3x^\prime
-{2 \over 3} \epsilon^{5/2} \stackrel{(3)\quad}{{\cal I}^{kk}(t)} \nonumber
\\
&&+ {1 \over 6} \epsilon^3 \partial^4_t \int_{\cal M}
\tau^{00}(t,{\bf x}^\prime) |{\bf x}-{\bf x}^\prime |^3 d^3x^\prime 
\nonumber \\
&&- {1 \over 30} \epsilon^{7/2} \left \{ (4x^{kl}+2r^2\delta^{kl})
\stackrel{(5)\quad}{{\cal I}^{kl}(t)}
- 4 x^k \stackrel{(5)\qquad}{{\cal I}^{kll}(t)}
+ \stackrel{(5)\qquad}{{\cal I}^{kkll}(t)} \right
\} \nonumber \\
&&+ {1 \over 180} \epsilon^4 \partial^6_t \int_{\cal M}
\tau^{00}(t,{\bf x}^\prime) |{\bf x}-{\bf x}^\prime |^5 d^3x^\prime 
\nonumber \\
&&- {1 \over 1260} \epsilon^{9/2} \left \{ 
3r^4 \stackrel{(7)\quad}{{\cal I}^{kk}(t)} 
+ 12r^2x^{ij} \stackrel{(7)\quad}{{\cal I}^{ij}(t)} 
-12 r^2 x^i \stackrel{(7)\qquad}{{\cal I}^{ikk}(t)}
\right .\nonumber \\
&& \left .
-8 x^{ijk}  \stackrel{(7)\qquad}{{\cal I}^{ijk}(t)}
+3r^2  \stackrel{(7)\qquad}{{\cal I}^{iikk}(t)}
+12 x^{ij}  \stackrel{(7)\qquad}{{\cal I}^{ijkk}(t)}
-6 x^i  \stackrel{(7)\qquad}{{\cal I}^{ikkll}(t)}
+  \stackrel{(7)\quad\quad }{{\cal I}^{iikkll}(t)} \right \}
\nonumber \\
&&+ N_{\partial {\cal M}} + O(\epsilon^5) \,, \\
%
%HERE IS K
%
K^i_{\cal N} &=& 4 \epsilon^{3/2} \int_{\cal M} {\tau^{0i}(t,{\bf x}^\prime) \over {|{\bf x}-{\bf
x}^\prime |}} d^3x^\prime +2 \epsilon^{5/2} \partial^2_t \int_{\cal M}
\tau^{0i}(t,{\bf x}^\prime) |{\bf x}-{\bf x}^\prime | d^3x^\prime
\nonumber \\
&& +{2 \over 9} \epsilon^3 \left \{ 3 x^k \stackrel{(4)\quad}{{\cal I}^{ik}(t)}
- \stackrel{(4)\quad}{{\cal I}^{ikk}(t)}
+2 \epsilon^{mik}  \stackrel{(3)\quad}{{\cal J}^{mk}(t)}
\right \} +  {1 \over 6} \epsilon^{7/2} \partial^4_t \int_{\cal M}
\tau^{0i}(t,{\bf x}^\prime) |{\bf x}-{\bf x}^\prime |^3 d^3x^\prime
\nonumber \\
&& + {1 \over 450} \epsilon^4 \left \{ 
30r^2 x^k \stackrel{(6)\quad}{{\cal I}^{ik}(t)}
- 10 r^2 \stackrel{(6)\qquad}{{\cal I}^{ikk}(t)}
-20 x^{kl} \stackrel{(6)\qquad}{{\cal I}^{ikl}(t)}
+15 x^k \stackrel{(6)\qquad}{{\cal I}^{ikll}(t)}
-3 \stackrel{(6)\qquad}{{\cal I}^{ikkll}(t)} \right .
\nonumber \\
&& \left . + \epsilon^{mil} \left [ 
20r^2 \stackrel{(5)\quad}{{\cal J}^{ml}(t)}
+40x^{kl} \stackrel{(5)\quad}{{\cal J}^{mk}(t)}
-15 x^l \stackrel{(5)\qquad}{{\cal J}^{mkk}(t)}
-30 x^k \stackrel{(5)\qquad}{{\cal J}^{mkl}(t)}
+12 \stackrel{(5)\qquad}{{\cal J}^{mlkk}(t)} \right ] \right \}
\nonumber \\
&&+ K^i_{\partial {\cal M}} + O(\epsilon^{9/2}) \,, \\
%
%HERE IS Bij
%
B^{ij}_{\cal N} &=& 4 \epsilon^2 \int_{\cal M} {\tau^{ij}(t,{\bf x}^\prime) 
\over {|{\bf x}-{\bf x}^\prime |}} d^3x^\prime 
- 2 \epsilon^{5/2} \stackrel{(3)\quad}{{\cal I}^{ij}(t)}
+2 \epsilon^3 \partial^2_t \int_{\cal M}
\tau^{ij}(t,{\bf x}^\prime) |{\bf x}-{\bf x}^\prime | d^3x^\prime
\nonumber \\
&& - {1 \over 9} \epsilon^{7/2} \left \{ 
3 r^2 \stackrel{(5)\quad}{{\cal I}^{ij}(t)}
-2x^k \stackrel{(5)\qquad}{{\cal I}^{ijk}(t)}
- 8 x^k \epsilon^{mk(i} \stackrel{(4)\qquad}{{\cal J}^{m|j)}(t)}
+ 6 \stackrel{(3)\qquad}{M^{ijkk}(t)} \right \}
\nonumber \\
&& + {1 \over 6} \epsilon^4 \partial^4_t \int_{\cal M}
\tau^{ij}(t,{\bf x}^\prime) |{\bf x}-{\bf x}^\prime |^3 d^3x^\prime
\nonumber \\
&&-{1 \over 180} \epsilon^{9/2} \left \{
3r^4 \stackrel{(7)\quad}{{\cal I}^{ij}(t)}
-4r^2 x^k \stackrel{(7)\qquad}{{\cal I}^{ijk}(t)}
-16r^2 x^k \epsilon^{mk(i} \stackrel{(6)\qquad}{{\cal J}^{m|j)}(t)}
\right . \nonumber \\
&& \left . 
+ 12r^2 \stackrel{(5)\qquad}{M^{ijkk}(t)}
+24 x^{kl} \stackrel{(5)\qquad}{M^{ijkl}(t)}
-24 x^k \stackrel{(5)\qquad}{M^{ijkll}(t)}
+6 \stackrel{(5)\qquad}{M^{ijkkll}(t)}  \right \} \nonumber \\
&&+ B^{ij}_{\partial {\cal M}} + O(\epsilon^5) \,.
\end{eqnarray}
\end{mathletters}
%label{bigexpansion}
Explicit formulae for the boundary terms $N_{\partial {\cal M}}$,
$K^i_{\partial {\cal M}}$ and $B^{ij}_{\partial {\cal M}}$ are given
in Appendix \ref{boundary}.  Through 3.5PN order, the terms in Eq.
(\ref{bigexpansion}) divide naturally into two types: {\it even}
terms, {\it i.e.} terms of integer powers in $\epsilon$ in $N$ and
$B^{ij}$ and odd-half integer powers in $K^i$, and {\it odd} terms, of
odd-half integer powers in $N$ and
$B^{ij}$ and integer powers in $K^i$.  The even terms produce the
leading Newtonian, PN, 2PN and 3PN contributions to the equations of
motion, while the odd terms produce the gravitational radiation
reaction forces.  (Note that the even terms have odd contributions
embedded within them, via contributions of the metric itself to
$\tau^{\alpha\beta}$).  Through 3.5PN order, there is a clean division
between even and odd terms, in the sense that even terms produce
non-dissipative contributions to the equations of motion, while odd
terms produce radiation reaction effects.  At 4PN order this
separation fails, because of the presence of tails -- these are
$O(\epsilon^{3/2})$ modifications of the leading 2.5PN
radiation-reaction terms, which result in disspative effects at 4PN
order.  We derive the leading contributions of these 4PN tail terms in
Sec. \ref{nearzonetails}.

The outer integrals for near-zone field points
turn out to contribute only beginning at 3PN order (and, as we will
see, do not contribute observable effects until 4PN order).  
This can be seen schematically
as follows:  for a source term of the form 
$f_{B,L}(u) {\hat n}^{<L>} r^{-B}$, the outer integral has the form
\begin{equation}
\int f(t-r^\prime ) {({\hat n}^\prime )}^{<L>}
(r^\prime)^{-B} {{d^3x^\prime} \over {|{\bf x}-{\bf x}^\prime |}}
\sim \int_{\cal R}^\infty {{d^q f(t)} \over dt^q} {r^\prime}^{q+1-B}
dr^\prime \sim {{d^q f(t)} \over dt^q} {\cal R}^{q+2-B} \,,
\label{quickanddirty}
\end{equation}
%label{quickanddirty}
where we have used the fact that $|{\bf x}| \ll |{\bf x}^\prime |$.
The only possible 
$\cal R$-independent terms come from the case $q=B-2$.  Thus the
outer integral gives a schematic contribution $h^{\alpha\beta}_{{\cal
C}-{\cal N}} \sim f^{(B-2)}(t)$ where the
superscript $(B-2)$ denotes $B-2$ time derivatives.
From Eq. (\ref{Lambda}a), the
leading contribution to the source comes from $(\nabla N)^2$,
where, from Eq. (\ref{genexpand}), 
$N$ has the far-zone form $N \approx 4{\cal I}/r + 2(3 {\hat n}^{<kl>}
{\cal I}^{kl}/r^3 + 3{\hat n}^{<kl>}{\dot {\cal I}}^{kl}/r^2 + {\hat n}^{kl}
{\ddot {\cal I}}^{kl}/r) +\dots$.  Taking the gradient of this expression and
squaring, we get, schematically $(\nabla N)^2 \sim {\cal I}^2/r^4
+ {\cal I}({\cal I}^{kl}/r^6 + {\dot {\cal I}}^{kl}/r^5 + {\ddot {\cal
I}}^{kl}/r^4 +\dots)$.  The first term ($B=4$) gives no
contribution, since ${\cal I}$ is constant to the order considered (${\cal
I}$ varies only via gravitational radiation energy loss).  The second,
third and fourth terms, ($B=6,\, 5 ,\, 4$) together give 
$h \sim {\cal I}{\cal I}^{kk(4)}(t)$.  Since ${\cal
I}^{kk} \sim mr^2$, we find $h \sim (m/r)^2 v^4 \sim O(\epsilon^4)$,
which is a 3PN contribution.  Thus, for near zone field points, the
outer integrals can be ignored until 3PN order.  A similar argument
for far-zone field points reveals that outer integrals begin to contribute 
only at 2PN order, as was found by WW \cite{opus}.

\subsection{Compendium of useful post-Newtonian near-zone potentials}
\label{compendium}

The even terms in Eq. (\ref{bigexpansion})
have the form of ordinary Poisson-like potentials and
their generalizations, sometimes called superpotentials.  For a source
$f$, we define the Poisson potential, superpotential, and
superduperpotential to be
\begin{mathletters}
\begin{eqnarray}
P(f) &\equiv& {1 \over {4\pi}} \int_{\cal M} {{f(t,{\bf x}^\prime)}
\over {|{\bf x}-{\bf x}^\prime | }} d^3x^\prime \,, \quad \nabla^2
P(f) = -f \,, \\
S(f)&\equiv& {1 \over {4\pi}} \int_{\cal M}f(t,{\bf x}^\prime)|{\bf
x}-{\bf x}^\prime | d^3x^\prime \,, \quad \nabla^2 S(f) = 2P(f) \,,
\\
SD(f)&\equiv&  {1 \over {4\pi}} \int_{\cal M}f(t,{\bf x}^\prime)|{\bf
x}-{\bf x}^\prime |^3 d^3x^\prime \,, \quad \nabla^2 SD(f) = 12S(f) \,.
\end{eqnarray}
\label{definepoisson}
\end{mathletters}
We also define potentials based on the ``densities'' $\sigma$,
$\sigma^i$ and $\sigma^{ij}$ constructed from $T^{\alpha\beta}$,
\begin{mathletters}
\begin{eqnarray}
\Sigma (f) &\equiv& \int_{\cal M} {{\sigma(t,{\bf x}^\prime)f(t,{\bf x}^\prime)}
\over {|{\bf x}-{\bf x}^\prime | }} d^3x^\prime = P(4\pi\sigma f) \,,
\\
\Sigma^i (f) &\equiv& \int_{\cal M} {{\sigma^i(t,{\bf x}^\prime)f(t,{\bf
x}^\prime)}
\over {|{\bf x}-{\bf x}^\prime | }} d^3x^\prime = P(4\pi\sigma^i f) \,,
\\
\Sigma^{ij} (f) &\equiv& \int_{\cal M} {{\sigma^{ij}(t,{\bf x}^\prime)f(t,{\bf
x}^\prime)}
\over {|{\bf x}-{\bf x}^\prime | }} d^3x^\prime = P(4\pi\sigma^{ij} f) \,,
\end{eqnarray}
\end{mathletters}
%label{definepoisson}
along with the superpotentials
\begin{mathletters}
\label{definesuper}
\begin{eqnarray}
X(f)  &\equiv& \int_{\cal M} {\sigma(t,{\bf x}^\prime)f(t,{\bf
x}^\prime)}
{|{\bf x}-{\bf x}^\prime | } d^3x^\prime = S(4\pi\sigma f) \,,
\\
Y(f) &\equiv& \int_{\cal M} {\sigma(t,{\bf x}^\prime)f(t,{\bf x}^\prime)}
{|{\bf x}-{\bf x}^\prime |^3 } d^3x^\prime = SD(4\pi\sigma f) \,,
\\
Z(f) &\equiv& \int_{\cal M} {\sigma(t,{\bf x}^\prime)f(t,{\bf x}^\prime)}
{|{\bf x}-{\bf x}^\prime |^5 } d^3x^\prime \,,
\end{eqnarray}
\end{mathletters}
%label{definesuper}
and their obvious counterparts $X^i$, $X^{ij}$, $Y^i$, $Y^{ij}$, and so on.
A number of potentials occur sufficiently frequently in the PN
expansion that is it useful to define them specifically.  First and
foremost is the ``Newtonian'' potential,
\begin{equation}
U \equiv \int_{\cal M} {{\sigma(t,{\bf x}^\prime)}
\over {|{\bf x}-{\bf x}^\prime | }} d^3x^\prime = P(4\pi\sigma) =
\Sigma(1) \,.
\\
\end{equation}
The potentials needed for the post-Newtonian limit are:
\begin{eqnarray}
V^i \equiv \Sigma^i(1) \,, \qquad  & \Phi_1^{ij} \equiv \Sigma^{ij}(1) \,,
\qquad & \Phi_1 \equiv \Sigma^{ii}(1) \,, \nonumber \\
\Phi_2 \equiv \Sigma(U) \,, \qquad & X \equiv X(1) \,.
\end{eqnarray}
Useful 2PN potentials include:
\begin{eqnarray}
&V_2^i \equiv \Sigma^i(U) \,, \qquad & \Phi_2^i \equiv \Sigma(V^i) \,,
\nonumber \\
&X^i \equiv X^i(1) \,, \qquad & X^{ij} \equiv X^{ij}(1) \,, \nonumber
\\
&X_1 \equiv  X^{ii} \,, \qquad & X_2 \equiv X(U) \,,\nonumber \\
&P_2^{ij} \equiv P(U^{,i}U^{,j}) \,, \qquad & P_2 \equiv P_2^{ii}=\Phi_2
-{1 \over 2}U^2 \,,\nonumber \\
&G_1 \equiv P({\dot U}^2)  \,, \qquad & G_2 \equiv P(U {\ddot U}) \,,
\nonumber \\
&G_3 \equiv -P({\dot U}^{,k} V^k) \,, \qquad & G_4 \equiv
P(V^{i,j}V^{j,i}) \,,\nonumber \\
&G_5 \equiv -P({\dot V}^k U^{,k}) \,, \qquad & G_6 \equiv P(U^{,ij}
\Phi_1^{ij}) \,,\nonumber \\
&G_7^i \equiv P(U^{,k}V^{k,i}) + {3 \over 4} P(U^{,i}\dot U ) \,,
\qquad & H \equiv P(U^{,ij} P_2^{ij}) \,. 
\label{potentiallist}
\end{eqnarray}
%label{potentiallist}
At 3PN order, the following potentials are useful:
\begin{eqnarray}
&Y_1 \equiv Y^{ii} \,, \qquad & Y_2 \equiv Y(U) \,, \qquad Z \equiv
Z(1) \,.
\end{eqnarray}

A variety of properties of these and general Poisson potentials are
described in Appendix \ref{poisson}. 
Note that, in evaluating Poisson potentials
and superpotentials of sources that do not have compact support, our
rule is to evaluate them on the finite, constant time hypersurface
$\cal M$, and to discard all terms that depend on $\cal R$.

\section{Expansion of near-zone fields to 2.5PN order}
\label{2.5expansion}

We now turn to explicit evaluation of the near-zone fields and the metric to
higher PN order, in terms of Poisson
potentials and multipole moments.  
In addition to evaluating the inner integrals shown
above, we must evaluate the outer integrals consistently at each PN
order, to check whether any finite, $\cal R$-independent 
contributions result.   

In evaluating the contributions at each order, we shall use the
following notation,
\begin{mathletters}
\label{expandNKB}
\begin{eqnarray}
N &=& \epsilon (N_0 + \epsilon N_1+ \epsilon^{3/2} N_{1.5}+ \epsilon^2 N_2+
\epsilon^{5/2} N_{2.5}+ \epsilon^3 N_3+ \epsilon^{7/2} N_{3.5} )
+O(\epsilon^5) \,, \\
K^i &=& \epsilon^{3/2} (K_1^i + \epsilon K_2^i +\epsilon^{3/2} K_{2.5}^i
+ \epsilon^2 K_3^i +\epsilon^{5/2} K_{3.5}^i ) +O(\epsilon^{9/2}) \,, \\
B &=& \epsilon^2 (B_1 + \epsilon^{1/2} B_{1.5} 
+\epsilon B_2 + \epsilon^{3/2} B_{2.5} + \epsilon^2 B_3+
\epsilon^{5/2} B_{3.5} )
+O(\epsilon^5) \,, \\
B^{ij} &=& \epsilon^2 (B_2^{ij} + \epsilon^{1/2} B_{2.5}^{ij} +
\epsilon B_3^{ij}+ \epsilon^{3/2} B_{3.5}^{ij}) +O(\epsilon^4) \,,
\end{eqnarray}
\end{mathletters}
%label{expandNKB}
where the subscript on each term indicates the level (1PN, 2PN, 2.5PN,
{\it etc.}) of its leading contribution to the equations of motion.  Notice
that our separate treatment of $B$ and $B^{ij}$ leads to the slightly
awkward notational circumstance that, for example, $B_2^{ii} = B_1$.

\subsection{The Newtonian and 1.5PN solution}

At lowest order in the PN expansion, we only need to evaluate
$\tau^{00} = (-g)T^{00} + O(\rho\epsilon) = \sigma  +
O(\rho\epsilon)$ (recall that $\sigma^{ii} \sim \epsilon \sigma$).
Since this has compact support, the outer integrals vanish, and we
find
\begin{equation}
N_0 = 4U \,.
\label{newtonian}
\end{equation}
%label{newtonian}
To this order, $(-g)=1+4U+O(\epsilon^2)$.  

To the next PN order, we obtain, from Eqs. (\ref{effective}),
(\ref{Lambda}) and  (\ref{newtonian}),
\begin{eqnarray}
\tau^{00} &=& \sigma - \sigma^{ii} + 4\sigma U - {7 \over
{8\pi}} \nabla U^2 + O(\rho\epsilon^2) \,, \nonumber \\
\tau^{0i} &=& \sigma^i + O(\rho\epsilon^{3/2}) \,, \nonumber \\
\tau^{ii} &=& \sigma^{ii} -{1 \over {8\pi}} \nabla U^2 + 
O(\rho\epsilon^2) \,, \nonumber \\
\tau^{ij} &=&  O(\rho\epsilon) \,. 
\label{tauPN}
\end{eqnarray}
%label{tauPN}
Substituting into Eqs. (\ref{bigexpansion}), and calculating terms
through 1.5PN order (e.g. $O(\epsilon^{5/2})$ in $N$), we obtain
\begin{mathletters}
\label{postnewtonian}
\begin{eqnarray}
N_1 &=& 7U^2-4\Phi_1+2\Phi_2+2{\ddot X} \,, \\
K_{1}^i &=& 4V^i \,,\\
B_1 &=& U^2+4\Phi_1-2\Phi_2 \,, \\
N_{1.5} &=& -{2 \over 3} \stackrel{(3)\quad}{{\cal I}^{kk}(t)}
\,,\\
B_{1.5} &=& -{2} \stackrel{(3)\quad}{{\cal I}^{kk}(t)} \,.
\end{eqnarray}
\end{mathletters}
%label{postnewtonian}
It is straightforward in this case to show that the outer integrals
and surface terms give no $\cal R$-independent terms.  It is
useful to illustrate the cancellation of an $\cal R$-dependent
term in this simple case.  In the far zone to Newtonian order, the
field, from Eq. (\ref{genexpand}), is given by $N \approx 4{\cal
I}/r$, where we focus on the monopole contribution.  This contributes
to $\Lambda^{00}$ in the far-zone a term of the form
$\Lambda^{00}=-14{\cal I}^2/r^4$.  To evaluate the near-zone
contribution of the outer integral of this term, we must evaluate the
coefficient ${\cal E}^q_{B,L}$ in Eq. (\ref{outernear})
with $q=0$ (no time derivative, since
$\cal I$ is constant, to lowest order), $B=4$, $L=0$.  From Eqs.
(\ref{outerdefinitions}) and (\ref{calE}), 
this yields a contribution to $N$ given by
$N_{{\cal C}-{\cal N}} = -7{\cal I}^2/{\cal R}^2$.  However, in evaluating
$N_{\cal N}$, we encounter the Poisson potential $-14P(\nabla U^2) =
-14P_2$ (see Eq. (\ref{potentiallist})).
Upon integrating by parts and keeping the surface term at $\cal R$
(see Eq. (\ref{Pformulae2}a)),
this gives a contribution $7U^2-14\Phi_2+7{\cal I}^2/{\cal R}^2$, 
whose $\cal R$-dependent term cancels
that from the outer integral.

The physical metric to 1.5PN order is then
\begin{mathletters}
\label{1.5pnmetric}
\begin{eqnarray}
g_{00} &=& -1 + 2U - 2U^2 + \ddot X 
- {4 \over 3}\stackrel{(3)\quad}{{\cal I}^{kk}(t)} 
+ O(\epsilon^3) \,,\\
g_{0i} &=& -4V^i + O(\epsilon^{5/2}) \,,\\
g_{ij} &=& \delta_{ij} (1+2U) + O(\epsilon^2) \,.
\end{eqnarray}
\end{mathletters}
%label{1.5pnmetric}

Notice that, in our formulation, the potential $U$ is {\it not} a
retarded potential; the retardation is expressed by the PN potential
$\ddot X$ and the 1.5PN term 
$-{4 \over 3} {{\cal I}^{kk(3)}(t)}$.
This contrasts with the PM approach, where retarded,
rather than Poisson potentials are used, and the retardation is
expanded only much later in the computation.
The apparently 1.5PN term 
$-{4 \over 3} {{\cal I}^{kk(3)}(t)}$
in $g_{00}$ actually doesn't contribute to the equations of motion
at this order because it is purely a function of time, and the
leading contribution is through a spatial gradient.  As a
consequence, the
lowest-order observable contribution to radiation reaction is at
2.5PN order.  (An alternative way to
treat this 1.5PN term would be to absorb it in a redefinition of the time
coordinate.)

We note here the useful identity, which follows from Eqs. (\ref{tauPN}), 
$\sigma = \tau^{00} +\tau^{ii} - (1/2\pi)\nabla^2 (U^2)
+ O(\rho \epsilon^2)$, whose consequence is,
\begin{equation}
\int_{\cal M} \sigma(t,{\bf x}) d^3x = {\cal I} + {1 \over 2} {\ddot
{\cal I}}^{ii} + O({\cal I} \epsilon^2) \,,
\label{massidentity}
\end{equation}
%label{massidentity}
where surface terms make no $\cal R$-independent contribution.

\subsection{The 2.5PN solution}

At 2PN and 2.5PN order, we obtain from 
Eqs. (\ref{effective}),
(\ref{Lambda}b,c), (\ref{newtonian}) and  (\ref{postnewtonian})
\begin{mathletters}
\begin{eqnarray}
\tau^{ij} &=& \sigma^{ij} + {1 \over 4\pi}(U^{,i}U^{,j} - {1 \over 2}
\delta^{ij} \nabla U^2 ) + O(\rho\epsilon^2) \,, \\
\tau^{0i} &=& \sigma^i + 4\sigma^i U+ {2 \over \pi} U^{,j}V^{[j,i]}
+ {3 \over 4\pi} \dot U U^{,i} + O(\rho\epsilon^{5/2}) \,.
\end{eqnarray}
\end{mathletters}
Including outer integrals and boundary terms (which contribute
nothing), we obtain, from Eq. (\ref{bigexpansion}c), 
\begin{mathletters}
\label{2postnewtonian}
\begin{eqnarray}
B_2^{ij} &=& 4\Phi_1^{ij} +4P_2^{ij}-\delta^{ij}(2\Phi_2-U^2)  \,,\\
K_2^i &=& 8V_2^i -8\Phi_2^i + 8UV^i + 16G_7^i + 2{\ddot X}^i
\,, \\
B_{2.5}^{ij} &=&
-2 \stackrel{(3)\quad}{{\cal I}^{ij}(t)}  \,, \\
K_{2.5}^i &=& {2 \over 3} x^k \stackrel{(4)\quad}{{\cal
I}^{ik}(t)}
- {2 \over 9} \stackrel{(4)\quad}{{\cal I}^{ikk}(t)}
+ {4 \over 9} \epsilon^{mik}  \stackrel{(3)\quad}{{\cal J}^{mk}(t)}
\,.
\end{eqnarray}
\end{mathletters}
%label{2postnewtonian}
The solutions 
for $B_2^{ij}$ and $B_{2.5}^{ij}$, along with the earlier 1.5PN solutions 
must now be substituted into $(-g)T^{\alpha\beta}$ and Eq.
(\ref{Lambda}a,d), with the result
\begin{mathletters}
\begin{eqnarray}
\tau^{00} &=& \sigma - \sigma^{ii} +4\sigma U- {7 \over 8\pi}\nabla U^2
\nonumber \\
&&+ \sigma ( 7U^2 -8\Phi_1+2\Phi_2 +2\ddot X )
-4\sigma^{ii} U\nonumber
\\
&& +{1 \over 4\pi} \left \{ {5 \over 2}{\dot U}^2 - 4U\ddot U -8{\dot
U}^{,k}V^k +2V^{i,j}(3V^{j,i}+V^{i,j}) +4{\dot V}^jU^{,j}
-4U^{,ij}\Phi_1^{ij} \right . 
\nonumber \\
&& 
\left . 
+8\nabla U \cdot \nabla
\Phi_1- 4\nabla U \cdot \nabla
\Phi_2 - {7 \over 2} \nabla U \cdot \nabla \ddot X
-10U\nabla U^2 -4U^{,ij} P_2^{ij}
\right \} 
\nonumber \\
&&
+ {4 \over 3} \sigma \stackrel{(3)\quad}{{\cal I}^{kk}(t)} 
+{1 \over {2\pi}} U^{,ij} \stackrel{(3)\quad}{{\cal I}^{ij}(t)} \,, \\
\tau^{ii} &=& \sigma^{ii} - {1 \over 8\pi}\nabla U^2
\nonumber \\
&& +4\sigma^{ii} U -{1 \over 4\pi} \left \{ {9 \over 2}{\dot U}^2 
+4V^{i,j}V^{[i,j]} +4{\dot V}^jU^{,j} 
+{1 \over 2} \nabla U \cdot \nabla \ddot X
\right \} \,.
\end{eqnarray}
\label{tau00ii2PN}
\end{mathletters}
%label{tau00ii2PN}
Substituting into Eq. (\ref{bigexpansion}a) and (\ref{bigexpansion}c) 
and evaluating terms
through $O(\epsilon^{7/2})$, and verifying that the outer integrals and
surface terms make no ${\cal R}$-independent contributions, we obtain,
\begin{mathletters}
\begin{eqnarray}
N_2 &=& 
-16U\Phi_1+8U\Phi_2+7U\ddot X+ {20 \over 3}U^3 -4V^iV^i 
-16\Sigma(\Phi_1)+\Sigma(\ddot X )+8\Sigma^i(V^i) \nonumber \\
&& -2{\ddot
X}_1+{\ddot X}_2 + {1 \over 6}\stackrel{(4)}{Y} 
-4G_1-16G_2+32G_3+24G_4-16G_5-16G_6-16H \,, \\
B_2 &=& U\ddot X + 4V^iV^i - \Sigma(\ddot X ) - 8\Sigma^i(V^i) 
+ 16\Sigma^{ii}(U) \nonumber \\
&&+ 2{\ddot X}_1-{\ddot X}_2  - 20G_1+8G_4 +16G_5 \,, \\
N_{2.5} &=& -{1 \over 15}(2x^{kl}+r^2\delta^{kl})\stackrel{(5)\quad}{{\cal
I}^{kl}(t)} + {2 \over 15} x^k\stackrel{(5)\qquad}{{\cal I}^{kll}(t)}
- {1 \over 30} \stackrel{(5)\qquad}{{\cal I}^{kkll}(t)}
\nonumber \\
&&+ {16 \over 3} U\stackrel{(3)\quad}{{\cal I}^{kk}(t)}
-4X^{,kl}\stackrel{(3)\quad}{{\cal I}^{kl}(t)} 
\,, \\
B_{2.5} &=&  
-{1 \over 3}  r^2 \stackrel{(5)\quad}{{\cal I}^{ii}(t)}
+{2 \over 9} x^k \stackrel{(5)\qquad}{{\cal I}^{iik}(t)}
+ {8 \over 9}  x^k \epsilon^{mki} \stackrel{(4)\qquad}{{\cal J}^{mi}(t)}
- {2 \over 3}  \stackrel{(3)\qquad}{M^{iikk}(t)} \,. 
\end{eqnarray}
\label{more2postnewtonian}
\end{mathletters}
%label{more2postnewtonian}

\subsection{The far-zone field to 1.5PN order}

In anticipation of finding non-zero 
outer-integral contributions to the near-zone field 
at 3PN order, we must determine the far-zone field to an order needed for 
the source $\Lambda^{\alpha\beta}$.    Our foregoing discussion indicates 
that counting PN orders for outer integrals is different than the standard 
method, because the inverse radial variable $r^{-1} < {\cal R}^{-1} \sim 
v/{\cal S}$; 
in other words, when considering contributions to the outer integrals, 
additional powers of $r$ in a term in the far-zone field can 
be regarded as increasing
the effective order of that term by half a power 
in $\epsilon$.   For example, expanding $h_{\cal N}^{00}=N _{\cal 
N}$ in the far-zone, Eq. (\ref{genexpand}), we obtain
\begin{eqnarray}
\label{Nfar}
N_{\cal N} &=& 4 \left \{ {{\cal I} \over r} + {1 \over 2} \partial_{kl} 
\left ( {{\cal I}^{kl}(u) \over r}  \right ) 
- {1 \over 6} \partial_{klm} \left ( 
{{\cal I}^{klm}(u) \over r}  \right ) + \dots \right \} \,, \\
&& \qquad \epsilon \qquad \qquad\quad \epsilon^2 \quad 
\qquad \qquad \epsilon^{5/2} \nonumber 
\end{eqnarray}
%label{Nfar}
where the effective PN order of each term is indicated.    In ordinary 
applications, the second potential in Eq. (\ref{Nfar})
would contribute a term of order $\epsilon$ 
of the form ${\hat n}^{<kl>} {\cal I}^{kl} /r^3$, which is simply the Newtonian 
quadrupole potential.  But in the outer integral, this term contributes an
$\cal R$-independent term only 
through several time derivatives, and thus its effective contribution
is higher order, in fact of the same order as that of the term ${\hat n}^{kl}
{\ddot{\cal I}}^{kl} /r$, which also comes from the second potential.

At this order, we must also be careful to include any outer integral 
and boundary 
contributions to the far-zone field.  From the lowest-order far-zone field, 
we find, to the order needed, that $\Lambda^{00} = -14({\cal I}/r^2)^2$, 
$\Lambda^{ij} = 4({\cal I}/r^2)^2 (n^{<ij>}-\delta^{ij}/6)$.  Evaluating 
the coefficients ${\cal D}^0_{4,2}$ and ${\cal D}^0_{4,0}$, Eq. (\ref{calD}), 
we obtain, in the far zone, $N_{{\cal C}-{\cal N}} = 7({\cal I}/r)^2$ 
and   $B_{{\cal C}-{\cal N}}^{ij} = ({\cal I}/r)^2{\hat n}^{ij}$.  
Combining the multipole expansions of Eq. (\ref{genexpand}) with the 
outer integral contributions, we obtain in the far-zone, to the order needed,
\begin{mathletters}
\label{farzone1.5PN}
\begin{eqnarray}
N &=& 4{{\cal I} \over r} 
+2 \partial_{kl} \left ( {{\cal I}^{kl}(u) \over r}  \right ) 
- {2 \over 3} \partial_{klm} \left ( {{\cal I}^{klm}(u) \over r}  \right ) 
+ 7 {{\cal I}^2 \over r^2} + O(\epsilon^3) \,, \\
K^i &=& -2 \partial_k  \left ( {{\dot{\cal I}}^{ik}(u) \over r}  \right ) 
+2\epsilon^{aib}{{{\hat n}^a {\cal J}^b} \over r^2} +{2 \over 3} 
\partial_{kl} \left ( {{\dot{\cal I}}^{ikl}(u) \over r}  \right ) +{4 \over 
3}\epsilon^{aib} \partial_{ak} \left ( {{\cal J}^{bk}(u) \over r}  \right ) + 
O(\epsilon^3) \,, \\
B^{ij} &=& 2 {{\ddot{\cal I}}^{ij}(u) \over r}+ {2 \over 3} \partial_{k} 
\left ( {{\ddot{\cal I}}^{ijk}(u) \over r}  \right ) +{8\over 3}\epsilon^{ak(i} 
\partial_{k} \left ( {{\dot{\cal J}}^{a|j)}(u) \over r}  \right ) 
+{{\cal I}^2 \over 
r^2}{\hat n}^{ij}+ O(\epsilon^3) \,.
\end{eqnarray}
\end{mathletters}
%label{farzone1.5PN}
It will turn out, however, that, despite the formal possibility of 3PN
contributions from the outer integrals, the {\it actual} contributions
will not begin until 4PN order (see Sec. {\ref{nearzonetails})

\section{Expansion of Near-Zone Fields to 3.5PN order}
\label{3.5expansion}

\subsection{$B^{ij}$ and $K^j$ to 3PN and 3.5PN order}

At 3PN and 3.5PN order, we 
obtain from
Eqs. (\ref{effective}),
(\ref{Lambda}b,c), (\ref{newtonian}) and  (\ref{postnewtonian})
\begin{mathletters}
\begin{eqnarray}
\tau^{ij} &=& \sigma^{ij} + {1 \over 4\pi}(U^{,i}U^{,j} - {1 \over 2}
\delta^{ij} \nabla U^2 ) \nonumber \\
&& + 4\sigma^{ij}U + {1 \over {4\pi}} \left \{ U^{,(i}{\ddot X}^{,j)} -16
V^{[i,k]}V^{[j,k]} + 8U^{,(i}{\dot V}^{j)} \right . \nonumber \\
&& \left . - \delta^{ij} ({1 \over 2} \nabla U \cdot \nabla {\ddot X}
-4V^{[l,k]}V^{[l,k]} + 4U^{,(k}{\dot V}^{k)} + {3 \over 2} {\dot U}^2
) \right \} + O(\rho\epsilon^3) \,, \\
\tau^{0i} &=& \sigma^i + 4\sigma^i U+ {2 \over \pi} U^{,j}V^{[j,i]}
+ {3 \over 4\pi} \dot U U^{,i} 
\nonumber \\
&& 
+\sigma^i(7U^2 -8\Phi_1+2\Phi_2+2 \ddot X ) 
\nonumber \\
&& 
+ {1 \over 16\pi} \left \{ 64U^{,k} (V_2^{[k,i]} - \Phi_2^{[k,i]} )
+32UU^{,k}V^{k,i} -16UU^{,k}V^{i,k} +16U^{,i}U^{,k}V^k \right . 
\nonumber \\
&& 
\left . -24V^i (\nabla U)^2 +16U^{,k}{\ddot X}^{[k,i]} +128U^{,k}G_7^{[k,i]}
-32 \Phi_1^{,k} V^{[k,i]} -16\Phi_2^{,k}V^{i,k} 
\right . 
\nonumber \\
&& 
\left . 
-16{\ddot X}^{,k}V^{[k,i]} 
-16\dot U \Phi_1^{,i} +48U \dot U U^{,i} +6\dot U {\ddot X}^{,i}
+6U^{,i} \stackrel{(3)}{X} -16U^{,i}{\dot \Phi}_1 +16 \dot U {\dot
V}^i 
\right . 
\nonumber \\
&& 
\left . 
- 16U{\ddot V}^i -32V^k {\dot V}^{i,k} -16V^{i,kl}( \Phi_1^{kl} + P_2^{kl} )
+16U^{,k} ({\dot \Phi}_1^{ik} +{\dot P}_2^{ik} ) \right . 
\nonumber \\
&& 
\left . 
+16 V^{k,l} (\Phi_1^{il,k} +\Phi_1^{ik,l}-\Phi_1^{kl,i})
+16 V^{k,l} (P_2^{il,k} +P_2^{ik,l}-P_2^{kl,i})
\right \} 
\nonumber
\\
&& 
+ {4 \over 3} \sigma^i\stackrel{(3)\quad}{{\cal I}^{kk}(t)} 
+ {1 \over 2\pi} \biggl ( V^{i,kl} \stackrel{(3)\quad}{{\cal I}^{kl}(t)} 
-U^{,k}  \stackrel{(4)\quad}{{\cal I}^{ik}(t)} \biggr ) \,,
\end{eqnarray}
\label{tauij0i3PN}
\end{mathletters}
%label{tauij0i3PN}
where the first line in each expression
is the contribution through 2PN order obtained earlier.
Substituting into Eq. (\ref{bigexpansion}b,c) and keeping contributions
through $O(\epsilon^{7/2})$, and checking that surface terms and outer
integrals make no contribution to this order, we obtain, 
\begin{mathletters}
\begin{eqnarray}
B_3^{ij} &=& 16\Sigma^{ij}(U) +4P( U^{,(i}{\ddot X}^{,j)}) -
64P(V^{[i,k]}V^{[j,k]}) +32P(U^{,(i}{\dot V}^{j)})+ 2{\ddot X}^{ij}
+2\ddot S (U^{,i}U^{,j}) 
\nonumber \\
&&
 +\delta^{ij} (U\ddot X - 4V^kV^k-\Sigma(\ddot X )
+8\Sigma^k(V^k)-{\ddot X}_2 -8G_1-8G_4+16G_5 ) \,, \\
K_3^i &=& 
12U^2V^i + 16UV_2^i -16U\Phi_2^i +4U{\ddot X}^i +32UG_7^i +4V^i{\ddot X} 
-8 \Phi_1 V^i + 8 \Phi_2 V^i 
\nonumber \\
&&
- 8V^k \Phi_1^{ik} - 8V^k P_2^{ik} 
-16\Sigma (V_2^i) +16\Sigma (\Phi_2^i) 
-16 \Sigma (UV^i) - 4\Sigma ({\ddot X}^i) -32\Sigma (G_7^i) 
\nonumber \\
&&
-24\Sigma^i
(\Phi_1) +4\Sigma^i({\ddot X}) +8 \Sigma^{kk}(V^i) 
+8\Sigma^k (\Phi_1^{ik}) +8\Sigma^k (P_2^{ik}) +8\Sigma^{ik} (V^k)
\nonumber \\
&&
+24P(U \dot U U^{,i}) +24 P(U^{,k} U^{,i} V^k ) 
+32P(U^{,k}V_2^{k,i}) - 32P(U^{,k}\Phi_2^{k,i})+ 64P(U^{,k}G_7^{k,i})
\nonumber \\
&& 
+8 P(U^{,k}{\ddot X}^{k,i}) +16P(U^{,k}{\dot \Phi}_1^{ik})
+16P(U^{,k}{\dot P}_2^{ik})
-16P(U^{,i}{\dot \Phi}_1) + 6P(U^{,i} \stackrel{(3)}{X}) 
\nonumber \\
&& 
-16P(\dot U \Phi_1^{,i} ) + 6P(\dot U {\ddot X}^{,i} )
+32P(UU^{,k}V^{i,k}) -16P(U{\ddot V}^i)
+ 16P(V^{k,l} \Phi_1^{il,k}) 
\nonumber \\
&& 
- 16P(V^{k,l} \Phi_1^{kl,i})
+ 16P(V^{k,l} P_2^{il,k}) 
-32P(V^k{\dot V}^{i,k}) -16P(V^{k,i} \Phi_1^{,k}) 
+8P(V^{k,i} {\ddot X}^{,k}) 
\nonumber \\
&&
- 16P(V^{k,l} P_2^{kl,i})
-16P(V^{i,lm} \Phi_1^{lm}) -16P(V^{i,lm} P_2^{lm})
\nonumber \\
&& 
+4 {\ddot X}^i (U) -4 {\ddot X} (V^i) +8 {\ddot S}(U^{,k}V^{k,i}) +
6  {\ddot S}(U^{,i}{\dot U}) + {1 \over 6} \stackrel{(4)}{Y^i} \,, \\
B_{3.5}^{ij} &=& 
-{1 \over 3} r^2 \stackrel{(5)\quad}{{\cal I}^{ij}(t)}
+{2 \over 9} x^k \stackrel{(5)\qquad}{{\cal I}^{ijk}(t)}
+ {8 \over 9} x^k \epsilon^{mk(i} \stackrel{(4)\qquad}{{\cal J}^{m|j)}(t)}
- {2 \over 3} \stackrel{(3)\qquad}{M^{ijkk}(t)} \,, \\
K_{3.5}^i &=& 
 {1 \over 15}r^2 x^k \stackrel{(6)\quad}{{\cal I}^{ik}(t)}
 - {1 \over 45} r^2 \stackrel{(6)\qquad}{{\cal I}^{ikk}(t)}
 -{2 \over 45} x^{kl} \stackrel{(6)\qquad}{{\cal I}^{ikl}(t)}
 +{1 \over 30} x^k \stackrel{(6)\qquad}{{\cal I}^{ikll}(t)}
 -{1 \over 150} \stackrel{(6)\qquad}{{\cal I}^{ikkll}(t)}
 \nonumber \\
 &&  + \epsilon^{mil} \left [
 {2 \over 45} r^2 \stackrel{(5)\quad}{{\cal J}^{ml}(t)}
 +{4 \over 45} x^{kl} \stackrel{(5)\quad}{{\cal J}^{mk}(t)}
 -{1 \over 30}  x^l \stackrel{(5)\qquad}{{\cal J}^{mkk}(t)}
 -{1 \over 15}  x^k \stackrel{(5)\qquad}{{\cal J}^{mkl}(t)}
 +{2 \over 75} \stackrel{(5)\qquad}{{\cal J}^{mlkk}(t)} \right ] 
 \nonumber \\
 && +{16 \over 3} V^i \stackrel{(3)\quad}{{\cal I}^{kk}(t)}
-4X^{i,kl} \stackrel{(3)\quad}{{\cal I}^{kl}(t)}
+4X^{,k} \stackrel{(4)\quad}{{\cal I}^{ik}(t)} \,.
\end{eqnarray}
\label{3postnewtonian}
\end{mathletters}
%label{3postnewtonian}

\subsection{$N$ and $B$ to 3PN and 3.5PN order}

The expressions for $\tau^{00}$ and $\tau^{ii}$ to 3PN and 3.5PN order are
too lengthy to be reproduced explicitly.  Instead, by substituting the
expansions (\ref{expandNKB}) into Eqs. (\ref{effective}) and
(\ref{Lambda}b,c), and keeping terms of $O(\rho\epsilon^3)$ and
$O(\rho\epsilon^{3.5})$, we obtain the formal contributions
\begin{mathletters}
\begin{eqnarray}
\tau_3^{00} &=&
\sigma(N_2-B_2-N_0B_1+K_1^iK_1^i)-\sigma^{ii}(N_1-B_1) 
\nonumber \\
&&
+{1 \over 16\pi} \biggl \{
-{7 \over 8} (2\nabla N_0 \cdot \nabla N_2 + (\nabla N_1 )^2 )
+{5 \over 4} {\dot N}_0{\dot N}_1 - {\ddot N}_0 N_1 - N_0{\ddot N}_1
\nonumber \\
&&
-2{\dot N}_0^{,i}K_2^i
-2{\dot N}_1^{,i}K_1^i
+K_1^{i,j}(3K_2^{j,i}+K_2^{i,j})
+N_0^{,i}{\dot K}_2^{i} +N_1^{,i}{\dot K}_1^{i}
-N_0^{,ij}B_3^{ij} 
\nonumber \\
&&
-N_1^{,ij}B_2^{ij}
+{1 \over 4}(\nabla N_0 \cdot \nabla B_2 +\nabla N_1 \cdot \nabla B_1)
+{7 \over 8}N_1 (\nabla N_0)^2  
\nonumber \\
&&
+{7 \over 4}N_0 (\nabla N_0 \cdot \nabla N_1)
+K_1^{i,j}{\dot B}_2^{ij} 
+{1 \over 4}B_2^{ij,k}(B_2^{ij,k}-2B_2^{jk,i})
+{1 \over 4}{\dot N}_0{\dot B}_1 -{1 \over 8}(\nabla B_1)^2
\nonumber \\
&&
+{1 \over 4}{\dot N}_0N_0^{,i}K_1^i
+{7 \over 8}N_0^{,i}N_0^{,j}B_2^{ij}
-{1 \over 2}N_0^{,i}K_1^j (4K_1^{i,j}+3K_1^{j,i})
-{7 \over 8}N_0^2 (\nabla N_0)^2 \biggr \}
\,, \\
\tau_{3.5}^{00} &=&
\sigma(N_{2.5}-B_{2.5}-N_0B_{1.5}) -\sigma^{ii}(N_{1.5}-B_{1.5})
\nonumber \\
&&
+{1 \over 16\pi} \biggl \{
-{7 \over 4}\nabla N_0 \cdot \nabla N_{2.5}
+{5 \over 4} {\dot N}_0{\dot N}_{1.5} 
- {\ddot N}_0 N_{1.5} - N_0{\ddot N}_{1.5}
-2{\dot N}_0^{,i}K_{2.5}^i
+N_0^{,i}{\dot K}_{2.5}^{i}
\nonumber \\
&&
+K_1^{i,j}(3K_{2.5}^{j,i}+K_{2.5}^{i,j})
-N_0^{,ij}B_{3.5}^{ij} -N_1^{,ij}B_{2.5}^{ij}
+{1 \over 4}\nabla N_0 \cdot \nabla B_{2.5} 
\nonumber \\
&&
+{7 \over 8}N_{1.5} (\nabla N_0)^2
+K_1^{i,j}{\dot B}_{2.5}^{ij}
+{1 \over 4}{\dot N}_0{\dot B}_{1.5}
+{7 \over 8}N_0^{,i}N_0^{,j}B_{2.5}^{ij} \biggr \} \,,
\\
\tau_3^{ii} &=&
\sigma^{ii} (N_1-B_1)
\nonumber \\
&&
+ {1 \over 16\pi} \biggl \{
-{1 \over 8} (2\nabla N_0 \cdot \nabla N_2 + (\nabla N_1)^2)
+2K_1^{i,j}K_2^{[j,i]}
-N_0^{,i}{\dot K}_2^i
-N_1^{,i}{\dot K}_1^i
\nonumber \\
&&
-{1 \over 4}(\nabla N_0 \cdot \nabla B_2 +\nabla N_1 \cdot \nabla B_1)
-{9 \over 4} {\dot N}_0 {\dot N}_1
+{1 \over 4}N_1 (\nabla N_0)^2
+{1 \over 2}N_0 (\nabla N_0 \cdot \nabla N_1)
\nonumber \\
&&
+2{\dot K}_1^i {\dot K}_1^i
-2 {\dot B}_1^{,i} K_1^i
+3 {\dot B}_2^{ij} K_1^{i,j}
-N_0 {\ddot B}_1
+{3 \over 4} {\dot N}_0 {\dot B}_1
+{1 \over 8} (\nabla B_1)^2
- B_1^{,ij}B_2^{ij}
\nonumber \\
&&
+{3 \over 4} B_2^{ij,k} B_2^{ij,k}
+{1 \over 2} B_2^{ij,k} B_2^{ik,j}
-N_0 K_1^{i,j} K_1^{[j,i]}
-{1 \over 2}N_0^{,i}K_1^j K_1^{j,i}
+N_0 N_0^{,i} {\dot K}_1^i
+{1 \over 4} {\dot N}_0 N_0^{,i} K_1^i
\nonumber \\
&&
-{1 \over 8} N_0^{,i} N_0^{,j} B_2^{ij}
+{1 \over 4} N_0 \nabla N_0 \cdot  \nabla B_1
+{1 \over 8} (\nabla N_0)^2 B_1
+{9 \over 8} N_0 {\dot N}_0^2
-{3 \over 8} N_0^2 (\nabla N_0)^2  \biggr \}
\,, \\
\tau_{3.5}^{ii} &=&
\sigma^{ii} (N_{1.5}-B_{1.5})
\nonumber \\
&&
+{1 \over 16\pi} \biggl \{ -{1 \over 4} \nabla N_0 \cdot \nabla N_{2.5}
+2 K_1^{i,j} K_{2.5}^{[j,i]}
-N_0^{,i} {\dot K}_{2.5}^i
-{1 \over 4} \nabla N_0 \cdot \nabla B_{2.5}
\nonumber \\
&&
- {9 \over 4} {\dot N}_0 {\dot N}_{1.5}
+ {1 \over 4} N_{1.5} (\nabla N_0)^2
+3 {\dot B}_{2.5}^{ij} K_1^{i,j}
-N_0 {\ddot B}_{1.5}
\nonumber \\
&&
+ {3 \over 4} {\dot N}_0 {\dot B}_{1.5}
-B_1^{,ij} B_{2.5}^{ij}
- {1 \over 8} N_0^{,i} N_0^{,j} B_{2.5}^{ij}
+ {1 \over 8} (\nabla N_0)^2B_{1.5} \biggr \}
\,.
\end{eqnarray}
\end{mathletters}
We have simplified the expressions slightly by taking into account the fact
that $N_{1.5}$, $B_{1.5}$ and $B_{2.5}^{ij}$ are purely functions of time,
so that spatial gradients of them vanish.  To obtain the full expressions,
one substitutes for $N_0$, $N_1$, $B_1$, $K_1^i$, {\it etc.} from Eqs.
(\ref{newtonian}), (\ref{postnewtonian}), (\ref{2postnewtonian}),
(\ref{more2postnewtonian}), and (\ref{3postnewtonian}).   Substituting this
into Eq. (\ref{bigexpansion}a) and (\ref{bigexpansion}c) (the latter
contracted on indices $ij$), and including surface terms and outer integrals,
we obtain the final 3PN and 3.5PN results for $N$ and
$B$:
\begin{mathletters}
\begin{eqnarray}
%
%HERE IS N_3
%
N_3 &=& 
{19 \over 6} U^4
-28U^2\Phi_1
+14U^2\Phi_2
+10U^2{\ddot  X}
-8U{\ddot X}_1
+4U{\ddot X}_2
+4U\Sigma({\ddot X}) 
\nonumber \\
&&
-4UG_1
-56UG_2
+112UG_3
+80UG_4
-64UG_5
-56UG_6
-56UH 
+{7 \over 12}U \stackrel{(4)}{Y}
\nonumber \\
&&
+32U\Sigma^i(V^i) 
-56U\Sigma(\Phi_1)
-8U\Sigma^{ii}(U)
+10\Phi_1^2
-8\Phi_1\Phi_2
+2\Phi_2^2
-8\Phi_1{\ddot  X}
+4\Phi_2 {\ddot  X}
\nonumber \\
&&
-16V^i V_2^i 
+ 16V^i \Phi_2^i 
-32 V^i G_7^i 
-4V^i {\ddot X}^i 
+ {7 \over 4} {\ddot X}^2
-2\Phi_1^{ij}\Phi_1^{ij}
-4\Phi_1^{ij} P_2^{ij} 
 -2 P_2^{ij} P_2^{ij}
\nonumber \\
&&
-8\Sigma(U\Phi_1) 
+36\Sigma(G_1) 
- 8\Sigma(G_2)
+16\Sigma(G_3) 
-48\Sigma(G_4) 
-8\Sigma(G_6) 
-8\Sigma(H)
\nonumber \\
&&
-8\Sigma({\ddot X}_1) 
+{1 \over 12}\Sigma(\stackrel{(4)}{Y})
+16\Sigma^i(V_2^i)  
-16\Sigma^i(\Phi_2^i) 
-16\Sigma^i(UV^i)
+32\Sigma^i(G_7^i)
+4\Sigma^i({\ddot X}^i)
\nonumber \\
&&
+8\Sigma^{ii}(U^2) 
+12\Sigma^{ii}(\Phi_1)
+4\Sigma^{ij}(\Phi_1^{ij}) 
+4\Sigma^{ij}(P_2^{ij})
-8\Sigma(\Sigma(\Phi_1)) 
+64\Sigma(\Sigma^i(V^i)) 
\nonumber \\
&&
-56\Sigma(\Sigma^{ii}(U))
-32P(U^2{\ddot U})
-28P(U{\dot U}^2)
+16P({\ddot U}\Phi_1)
+16P({\dot U}{\dot \Phi}_1)
\nonumber \\
&&
+16P(U{\ddot \Phi}_1)
-8P(U{\ddot \Phi}_2)
-8P(U \stackrel{(4)}{X})
-4P({\dot U} \stackrel{(3)}{X})
-8P({\ddot U}{\ddot X})
+32P(UU^{,i}{\dot V}^i)
\nonumber \\ 
&&
-64P(U{\dot U}^{,i} V^i)
-40P({\dot U}U^{,i}V^i)
+32P(UV^{i,j} V^{i,j})
+16P(UV^{i,j}V^{j,i})
\nonumber \\ 
&&
+32P(U^{,i}{\dot V}_2^i)
-32P(U^{,i}{\dot \Phi}_2^i)
+64P(U^{,i} {\dot G}_7^i)
+8P(U^{,i} \stackrel{(3)}{X^i})
\nonumber \\
&&
+64P({\dot U}^{,i} \Phi_2^i)
-64P({\dot U}^{,i} V_2^i)
-128P({\dot U}^{,i} G_7^i)
-16P({\dot U}^{,i}{\ddot X}^i)
\nonumber \\
&&
+4P(U^{,i}U^{,j}\Phi_1^{ij})
+4P(U^{,i}U^{,j}P_2^{ij})
-64P(U^{,ij}\Sigma^{ij}(U))
-8P(U^{,ij} {\ddot S}(U^{,i}U^{,j})) 
\nonumber \\
&&
-128P(U^{,ij}P(U^{,(i}{\dot V}^{j)})) 
-16P(U^{,ij}P(U^{,(i}{\ddot X}^{,j)})) 
+256P(U^{,ij} P(V^{[k,i]}V^{[k,j]}))
\nonumber \\
&&
-8P(U^{,ij}{\ddot X}^{ij}) 
+32P(V^i {\dot \Phi}_1^{,i})
-16P(V^i {\dot \Phi}_2^{,i})
-16P(V^i \stackrel{(3)}{X^{,i}})
+16P(V^{i,j}{\dot \Phi}_1^{ij})
\nonumber \\
&&
+16P(V^{i,j}{\dot P}_2^{ij})
+96P(V^{i,j}V_2^{j,i})
-96P(V^{i,j}\Phi_2^{j,i})
+192P(V^{i,j} G_7^{j,i})
\nonumber \\
&&
+24P(V^{i,j}{\ddot X}^{j,i})
+8P({\dot V}^i{\dot V}^i)
-16P({\dot V}^i \Phi_1^{,i})
+8P({\dot V^i}{\ddot X}^{,i})
-8P(\Phi_1^{ij,k} \Phi_1^{jk,i})
\nonumber \\
&&
-16P(\Phi_1^{ij,k} P_2^{jk,i})
-8P(P_2^{ij,k} P_2^{jk,i})
+16P(\Phi_1^{,ij} \Phi_1^{ij})
+16P(\Phi_1^{,ij} P_2^{ij})
-8P(\Phi_1^{ij} {\ddot X}^{,ij})
\nonumber \\
&&
-8P(\Phi_2^{,ij} \Phi_1^{ij})
-8P(\Phi_2^{,ij} P_2^{ij})
-8P(P_2^{ij} {\ddot X}^{,ij})
-{1 \over 6} \stackrel{(4)}{Y_1} 
+{1 \over 12} \stackrel{(4)}{Y_2} 
+{1 \over 180}\stackrel{(6)}{Z}
\nonumber \\
&&
-8{\ddot X}(\Phi_1)
+{1 \over 2}{\ddot X}({\ddot X})
+4{\ddot X}^i(V^i)
-2{\ddot S}({\dot U}^2)
-8{\ddot S}(U{\ddot U})
\nonumber \\
&&
-8{\ddot S}(U^{,ij}\Phi_1^{ij})
-8{\ddot S}(U^{,ij}P_2^{ij})
+12{\ddot S}(V^{i,j}V^{j,i})
+8{\ddot S}(U^{,i}{\dot V}^i)
-16{\ddot S}(V^{i}{\dot U}^{,i})
\,, \\
%
%HERE IS B_3
%
B_3 &=& 
{1 \over 2} U^4
 -4U^2\Phi_1
-2U^2\Phi_2 
+16UV^iV^i
-12UG_1
-8UG_2
+16UG_3
+16UG_4
\nonumber \\
&&
-8UG_6
-8UH 
+{1 \over 12}U \stackrel{(4)}{Y}
-8U\Sigma(\Phi_1)
+8U\Sigma^{ii}(U)
-2\Phi_1^2
+8\Phi_1\Phi_2
+2\Phi_2^2
\nonumber \\
&&
+16V^i V_2^i 
-16V^i \Phi_2^i 
+32 V^i G_7^i 
+4V^i {\ddot X}^i 
+ {1 \over 4} {\ddot X}^2
-6\Phi_1^{ij}\Phi_1^{ij}
-12 \Phi_1^{ij} P_2^{ij}
 -6 P_2^{ij} P_2^{ij}
\nonumber \\
&&
+8\Sigma(U\Phi_1) 
+12\Sigma(G_1) 
+8\Sigma(G_2)
-16\Sigma(G_3) 
-16\Sigma(G_4) 
+8\Sigma(G_6) 
+8\Sigma(H)
\nonumber \\
&&
-{1 \over 12}\Sigma(\stackrel{(4)}{Y})
-16\Sigma^i(V_2^i) 
+16\Sigma^i(\Phi_2^i) 
-48\Sigma^i(UV^i)
-32\Sigma^i(G_7^i)
-4\Sigma^i({\ddot X}^i)
+24\Sigma^{ii}(U^2) 
\nonumber \\
&&
-28\Sigma^{ii}(\Phi_1)
+8\Sigma^{ii}({\ddot X})
+12\Sigma^{ij}(\Phi_1^{ij}) 
+12\Sigma^{ij}(P_2^{ij})
+8\Sigma(\Sigma(\Phi_1)) 
-8\Sigma(\Sigma^{ii}(U)) 
\nonumber \\
&&
-68P(U{\dot U}^2)
-16P(U{\ddot \Phi}_1)
+8P(U{\ddot \Phi}_2)
+48P({\dot U}{\dot \Phi}_1)
-20P({\dot U} \stackrel{(3)}{X})
-56P({\dot U}U^{,i}V^i)
\nonumber \\
&&
-32P(UU^{,i}{\dot V}^i)
+32P(UV^{i,j} V^{i,j})
-16P(UV^{i,j}V^{j,i})
-32P(U^{,i}{\dot V}_2^i)
+32P(U^{,i}{\dot \Phi}_2^i)
\nonumber \\
&&
-64P(U^{,i} {\dot G}_7^i)
-8P(U^{,i} \stackrel{(3)}{X^i})
-4P(U^{,i}U^{,j}\Phi_1^{ij})
-4P(U^{,i}U^{,j}P_2^{ij})
-32P(V^i {\dot \Phi}_1^{,i})
\nonumber \\
&&
+16P(V^i {\dot \Phi}_2^{,i})
+48P(V^{i,j}{\dot \Phi}_1^{ij})
+48P(V^{i,j}{\dot P}_2^{ij})
+32P(V^{i,j}V_2^{j,i})
-32P(V^{i,j}\Phi_2^{j,i})
\nonumber \\
&&
+64P(V^{i,j} G_7^{j,i})
+8P(V^{i,j}{\ddot X}^{j,i})
+24P({\dot V}^i{\dot V}^i)
+16P({\dot V}^i \Phi_1^{,i})
-8P({\dot V^i}{\ddot X}^{,i})
\nonumber \\
&&
+8P(\Phi_1^{ij,k} \Phi_1^{jk,i})
+16P(\Phi_1^{ij,k} P_2^{jk,i})
+8P(P_2^{ij,k} P_2^{jk,i})
-16P(\Phi_1^{,ij} \Phi_1^{ij})
-16P(\Phi_1^{,ij} P_2^{ij})
\nonumber \\
&&
+8P(\Phi_2^{,ij} \Phi_1^{ij})
+8P(\Phi_2^{,ij} P_2^{ij})
+{1 \over 6} \stackrel{(4)}{Y_1} 
-{1 \over 12} \stackrel{(4)}{Y_2} 
-{1 \over 2}{\ddot X}({\ddot X})
-4{\ddot X}^i(V^i)
+8{\ddot X}^{ii}(U)
\nonumber \\
&&
-10{\ddot S}({\dot U}^2)
+4{\ddot S}(V^{i,j}V^{j,i})
-8{\ddot S}(U^{,i}{\dot V}^i)
+{4 \over 3}{\cal I}\stackrel{(4)\quad}{{\cal I}^{jj}(t)}  
\,, \\
%
%HERE IS N_3.5
%
N_{3.5} &=&
-{1 \over 420}r^4 \stackrel{(7)\quad}{{\cal I}^{jj}(t)}
-{1 \over 105} r^2 x^{ij}\stackrel{(7)\quad}{{\cal I}^{ij}(t)}
+{1 \over 105} r^2x^i \stackrel{(7)\qquad}{{\cal I}^{ijj}(t)}
+{2 \over 315} x^{ijk} \stackrel{(7)\qquad}{{\cal I}^{ijk}(t)}
\nonumber \\
&&
-{1 \over 420}r^2  \stackrel{(7)\qquad}{{\cal I}^{iijj}(t)}
-{1 \over 105} x^{ij} \stackrel{(7)\qquad}{{\cal I}^{ijkk}(t)}
+{1 \over 210} x^i \stackrel{(7)\qquad}{{\cal I}^{ijjkk}(t)}
-{1 \over 1260} \stackrel{(7)\quad \quad}{{\cal I}^{iijjkk}(t)}
\nonumber \\
&&
-\left ({8 \over 15}x^{ij}U +{6 \over 5}x^iX^{,j}+{2 \over 3}r^2X^{,ij}
-{2 \over 9}x^kY^{,ijk}+{11 \over 45}Y^{,ij} \right )
\stackrel{(5)\quad}{{\cal I}^{ij}(t)}
\nonumber \\
&&
+ \left ({16 \over 15}r^2U-{34 \over 15}x^iX^{,i}+{16 \over 5}X \right )
\stackrel{(5)\quad}{{\cal I}^{jj}(t)}
- {1 \over 45} (16x^iU-52X^{,i})\stackrel{(5)\qquad}{{\cal I}^{ijj}(t)}
\nonumber \\
&&
+ \left ( {4 \over 9} x^kX^{,ij}- {2 \over 27}Y^{,ijk} \right )
\stackrel{(5)\qquad}{{\cal I}^{ijk}(t)}
-{2 \over 15}U\stackrel{(5)\qquad}{{\cal I}^{iijj}(t)}
\nonumber \\
&&
- \left ( {4 \over 3}X^{i,j}- {8 \over 3}x^i{\dot X}^{,j}+ {10 \over 9}{\dot
Y}^{,ij} \right ) \stackrel{(4)\quad}{{\cal I}^{ij}(t)}
+8{\dot X} \stackrel{(4)\quad}{{\cal I}^{jj}(t)}
-{8 \over 9} {\dot X}^{,i} \stackrel{(4)\qquad}{{\cal I}^{ijj}(t)}
\nonumber \\
&&
- \left ( 14UX^{,ij}+2\Sigma(X^{,ij})-4X_1^{,ij}+2X_2^{,ij} + {2 \over
3}{\ddot Y}^{,ij} \right ) \stackrel{(3)\quad}{{\cal I}^{ij}(t)}
\nonumber \\
&&
+ {1 \over 3} (70U^2-16\Phi_1+20\Phi_2+4{\ddot X})\stackrel{(3)\quad}{{\cal
I}^{jj}(t)}
+{8 \over 3}U\stackrel{(3)\qquad}{{M}^{jjkk}(t)}
-{4 \over 3}X^{,ij}\stackrel{(3)\qquad}{{M}^{ijkk}(t)} 
\nonumber \\
&&
+{16 \over 9}x^kX^{,ij}\epsilon^{mk(i} \stackrel{(4)\quad}{{\cal J}^{m|j)}(t)}
-{4 \over 9} (8x^kU-5X^{,k})\epsilon^{mkj}\stackrel{(4)\quad}{{\cal J}^{mj}(t)}
+{16 \over 9}{\dot X}^{,k}\epsilon^{mkj}\stackrel{(3)\quad}{{\cal J}^{mj}(t)}
\,, \\
%
%HERE IS B_3.5
%
B_{3.5} &=&
-{1 \over 60}r^4 \stackrel{(7)\quad}{{\cal I}^{jj}(t)}
+{1 \over 45} r^2x^i \stackrel{(7)\qquad}{{\cal I}^{ijj}(t)}
-{1 \over 5} (Y^{,ij}-6x^iX^{,j})\stackrel{(5)\quad}{{\cal I}^{ij}(t)}
-{1 \over 5} (16X+2x^iX^{,i})\stackrel{(5)\quad}{{\cal I}^{jj}(t)}
\nonumber \\
&&
-{4 \over 15} X^{,i}\stackrel{(5)\qquad}{{\cal I}^{ijj}(t)}
+12X^{i,j}\stackrel{(4)\quad}{{\cal I}^{ij}(t)}
+ \left ( 6U^2+ {16 \over 3} \Phi_1-12\Phi_2 \right )
\stackrel{(3)\quad}{{\cal I}^{jj}(t)}
\nonumber \\
&&
- (2UX^{,ij}-2\Sigma(X^{,ij})+4X_1^{,ij}-2X_2^{,ij}-8P_2^{ij} )
\stackrel{(3)\quad}{{\cal I}^{ij}(t)}
\nonumber \\
&&
+{4 \over 45}r^2x^k \epsilon^{mkj}\stackrel{(6)\quad}{{\cal J}^{mj}(t)}
+{4 \over 3} X^{,k} \epsilon^{mkj}\stackrel{(4)\quad}{{\cal J}^{mj}(t)}
\nonumber \\
&&
-{1 \over 15} r^2 \stackrel{(5)\qquad}{{M}^{jjkk}(t)}
-{2 \over 15} x^{jk}\stackrel{(5)\qquad}{{M}^{iijk}(t)}
+{2 \over 15} x^{j}\stackrel{(5)\qquad}{{M}^{iijkk}(t)}
-{1 \over 30} \stackrel{(5)\qquad \quad}{{M}^{iijjkk}(t)} \,.
\end{eqnarray}
\label{NBfinal}
\end{mathletters}
%label{NBfinal}
The final term in the expression for $B_3$
is purely a function of time, and as such does not affect the equations
of motion through 3.5PN order.  It comes in part from the surface terms 
Eqs. (\ref{surfaceterms}), and in 
part from various integrations by parts of Poisson
potentials to achieve the expressions shown.  
In $N_3$, all such terms cancel.  Similarly, purely time-dependent
terms which appear in $N_{3.5}$ and $B_{3.5}$ do not contribute to the
equations of motion.  

As
expected, the outer integrals make their first formal contribution to
the field at 3PN order, however, the {\it observable} contribution vanishes
to this order, so we have not shown any such contributions 
explicitly
in Eqs. (\ref{NBfinal}).  In the next subsection, we study the contributions
of the outer integrals
in more detail, and show that through 3.5PN order, all contributions
from the outer integrals are pure gauge terms.

\subsection{Outer integrals and the contributions of ``tails''}
\label{nearzonetails}

Our earlier qualitative discussion suggested that 
terms involving products of the monopole moment $\cal I$
and the quadrupole moment ${\cal I}^{ij}$ of the far-zone fields would
contribute via the outer integrals at 3PN order.
Because higher multipole moments involve higher powers of $1/r$ or
higher time derivatives, they would be expected to contribute at
even higher PN order.  Thus working through 3.5PN order, we might expect
at most
that 
products of $\cal I$ with quadrupole ${\cal I}^{ij}$, octupole ${\cal I}^{ijk}$ or current quadrupole ${\cal J}^{ij}$
moments would contribute.  Other terms, such as products of $\cal I$ with
higher-order moments, or products of higher-order moments, such as
terms quadratic in ${\cal I}^{ij}$, will be
4PN order or higher.  In studying the contribution of the outer
integrals to the fields at 3.5PN order, therefore, it suffices
to employ the far-zone field given in Eq. (\ref{farzone1.5PN}).
However, to illustrate the first non-trivial ``tail''
contribution, we will evaluate certain pieces of the 
outer integrals through 4PN order.

We substitute Eqs. (\ref{farzone1.5PN})
into Eqs. (\ref{Lambda}) using the
``quick and dirty'' rule expressed by Eq. (\ref{quickanddirty}) to
determine which terms to keep, and 
obtain, in the far zone,
\begin{mathletters}
\begin{eqnarray}
\Lambda^{00} &=& 
14{\cal I}r^{-2}{\hat n}^i \partial_{ijk} ({\cal I}^{jk}/r)
-8{\cal I}r^{-1}\partial_{ij} ({\ddot {\cal I}}^{ij}/r)
+8 {\cal I}r^{-2}{\hat n}^i \partial_{j} ({\ddot {\cal I}}^{ij}/r)
-24{\cal I}r^{-4}{\hat n}^{<ij>} {\ddot {\cal I}}^{ij}
\nonumber \\ 
&&
-2{\cal I}r^{-2}{\hat n}^i \partial_{i} ({\ddot {\cal I}}^{jj}/r) 
-{14 \over 3} {\cal I}r^{-2}{\hat n}^i \partial_{ijkl} ({\cal
I}^{jkl}/r)
+{8 \over 3}{\cal I}r^{-1}\partial_{ijk} ({\ddot {\cal I}}^{ijk}/r)
\nonumber \\
&&
-{8 \over 3}{\cal I}r^{-2}{\hat n}^i \partial_{jk} ({\ddot {\cal I}}^{ijk}/r)
-8 {\cal I}r^{-3}{\hat n}^{<ij>} \partial_{k} ({\ddot {\cal I}}^{ijk}/r)
-{2 \over 3}{\cal I}r^{-2}{\hat n}^i \partial_{ij} ({\ddot {\cal
I}}^{jkk}/r)
\nonumber \\
&&
+{16 \over 3} {\cal I}r^{-2}{\hat n}^i \epsilon^{aib} \partial_{ak}
({\dot {\cal J}}^{bk}/r)
-32 {\cal I}r^{-3}{\hat n}^{<ij>}\epsilon^{aki} \partial_{k}({\dot
{\cal J}}^{aj}/r)
-{8 \over 3} {\cal I}r^{-2}{\hat n}^i \epsilon^{akj} \partial_{ik}
({\dot {\cal J}}^{aj}/r) 
\nonumber \\
&&
+ O(\rho \epsilon^4) \,,\\
\Lambda^{0i} &=& 
8{\cal I}r^{-2}{\hat n}^j \partial_{ik} ({\dot {\cal I}}^{jk}/r)
-8{\cal I}r^{-2}{\hat n}^j \partial_{jk} ({\dot {\cal I}}^{ik}/r)
-6{\cal I}r^{-2}{\hat n}^i \partial_{jk} ({\dot {\cal I}}^{jk}/r)
\nonumber \\
&&
+8{\cal I}r^{-1} \partial_{j} (\stackrel{(3)}{{\cal I}^{ij}}/r)
-{8 \over 3} {\cal I}r^{-1} \partial_{jk} (\stackrel{(3)}{{\cal
I}^{ijk}}/r)
+{16 \over 3} {\cal I}r^{-1} \epsilon^{iab} \partial_{ak}
({\ddot {\cal J}}^{bk}/r)
\nonumber \\
&&
-8 {\cal I}r^{-2}{\hat n}^j (\stackrel{(3)}{{\cal I}^{ij}}/r)
+2{\cal I}r^{-2}{\hat n}^i (\stackrel{(3)}{{\cal I}^{jj}}/r)
+O(\rho \epsilon^{7/2}) \,, \\
\Lambda^{ii} &=&
2 {\cal I}r^{-2}{\hat n}^i \partial_{ijk} ({\cal I}^{jk}/r)
-8 {\cal I}r^{-2}{\hat n}^i \partial_{j} ({\ddot {\cal I}}^{ij}/r)
+2{\cal I}r^{-2}{\hat n}^i \partial_{i} ({\ddot {\cal I}}^{jj}/r)
\nonumber \\
&&
-8{\cal I}r^{-2} \stackrel{(4)}{{\cal I}^{ii}}
-{2 \over 3}{\cal I}r^{-2}{\hat n}^i \partial_{ijkl} ({\cal I}^{jkl}/r)
+{8 \over 3}{\cal I}r^{-2}{\hat n}^i \partial_{jk} ({\ddot {\cal
I}}^{ijk}/r)
\nonumber \\
&&
+{2 \over 3}{\cal I}r^{-2}{\hat n}^i \partial_{ij} ({\ddot {\cal
I}}^{jkk}/r)
-{8 \over 3}{\cal I}r^{-1} \partial_{i} (\stackrel{(4)}{{\cal
I}^{ijj}}/r)
-{16 \over 3} {\cal I}r^{-2}{\hat n}^i \epsilon^{aib} \partial_{ak}
({\dot {\cal J}}^{bk}/r)
\nonumber
\\ &&
+{8 \over 3} {\cal I}r^{-2}{\hat n}^i \epsilon^{akj} \partial_{ik}
({\dot {\cal J}}^{aj}/r)
-{32 \over 3}{\cal I}r^{-1}\epsilon^{aki} \partial_{k}(\stackrel{(3)}{{\cal
J}^{ai}}/r)
+ O(\rho \epsilon^4) \,,
\\
\Lambda^{ij} &=& 
-8{\cal I}r^{-2} \stackrel{(4)}{{\cal I}^{ij}} + O(\rho \epsilon^3) \,.
\end{eqnarray}
\end{mathletters}
All moments ${\cal I}^{ij}$, ${\cal I}^{ijk}$, and ${\cal J}^{ij}$
in these expression are functions of retarded time $t-r$.  
Notice that the term kept in $\Lambda^{ij}$ is actually of $O(\rho
\epsilon^3)$  (4PN order) according to our scheme, however because it 
has $1/r^2$
dependence, it will yield a 4PN tail contribution of a form 
which we wish to keep.

We expand the derivatives and evaluate the coefficients ${\cal
E}_{B,L}^q$ and ${\cal E}_{2,L}^q$ [Eqs. (\ref{calE}) and (\ref{calE2})]
for each term, throwing away all
$\cal R$-dependent terms.  Terms with $1/r^2$ fall-off yield integrals
over Legendre functions $Q_L$, as in Eq. (\ref{outernear2}).  
The
result, through 3.5PN order (and keeping all formally 4PN terms involving
integrals over $Q_L$),  is
\begin{mathletters}
\begin{eqnarray}
(N_3)_{{\cal C}-{\cal N}} &=& 
{\cal I} \biggl \{ -8 {\hat n}^{<ij>} \int_1^\infty \stackrel{(4)}{{\cal
I}^{ij}}(t-r\zeta) Q_2(\zeta)d\zeta
-{8 \over 3} \int_1^\infty \stackrel{(4)}{{\cal
I}^{jj}}(t-r\zeta) Q_0(\zeta)d\zeta
\nonumber \\
&&
 \quad +{4 \over 3} ({\hat n}^{<ij>} - 2\delta^{ij} -2\delta^{ij}\ln r)
\stackrel{(4)}{{\cal I}^{ij}}(t) \biggr \} 
\,, \\
(N_{3.5})_{{\cal C}-{\cal N}} &=&
{\cal I} \biggl \{ 
-{8 \over 3}{\hat n}^{<ijk>} \int_1^\infty \stackrel{(5)}{{\cal
I}^{ijk}}(t-r\zeta) Q_3(\zeta)d\zeta
-{8 \over 5}{\hat n}^{i} \int_1^\infty \stackrel{(5)}{{\cal
I}^{ijj}}(t-r\zeta) Q_1(\zeta)d\zeta
\nonumber \\
&&
\quad -{2 \over 3} r(3{\hat n}^{<ij>}-2\delta^{ij})
\stackrel{(5)}{{\cal I}^{ij}}(t)
+{2 \over 45} (5{\hat n}^{<ijk>} + 18{\hat n}^{i}\delta^{jk})
\stackrel{(5)}{{\cal I}^{ijk}}(t) \biggr \}
\,, \\
%(N_{4})_{{\cal C}-{\cal N}} &=&
%{\cal I} \biggl \{ 
%-{2 \over 3} {\hat n}^{<ijkl>} \int_1^\infty \stackrel{(6)}{{\cal
%I}^{ijkl}}(t-r\zeta) Q_4(\zeta)d\zeta
%-{4 \over 3} {\hat n}^{<ij>} \int_1^\infty \stackrel{(6)}{{\cal
%I}^{ijkk}}(t-r\zeta) Q_2(\zeta)d\zeta
%-{2 \over 15} \int_1^\infty \stackrel{(6)}{{\cal
%I}^{iikk}}(t-r\zeta) Q_0(\zeta)d\zeta \biggr \}
%\,, \\
(K_{3.5})^i_{{\cal C}-{\cal N}} &=&
{\cal I} \biggl \{ -8{\hat n}^{j}\int_1^\infty \stackrel{(4)}{{\cal
I}^{ij}}(t-r\zeta) Q_1(\zeta)d\zeta
+ 4{\hat n}^j \stackrel{(4)}{{\cal I}^{ij}}(t) \biggr \}
\,,\\
(K_4)^i_{{\cal C}-{\cal N}} &=&
{\cal I} \biggl \{
-{8 \over 3} {\hat n}^{<jk>} \int_1^\infty \stackrel{(5)}{{\cal
I}^{ijk}}(t-r\zeta) Q_2(\zeta)d\zeta
-{8 \over 9}  \int_1^\infty \stackrel{(5)}{{\cal
I}^{ijj}}(t-r\zeta) Q_0(\zeta)d\zeta
\nonumber \\
&&
\quad +{16 \over 3} {\hat n}^{<ak>} \epsilon^{iaj} \int_1^\infty \stackrel{(4)}{{\cal
J}^{jk}}(t-r\zeta) Q_2(\zeta)d\zeta
\nonumber \\
&&
\quad +{16 \over 9} \epsilon^{ikj} \int_1^\infty \stackrel{(4)}{{\cal
J}^{jk}}(t-r\zeta) Q_0(\zeta)d\zeta \biggr \}
\,, \\
(B_3)_{{\cal C}-{\cal N}} &=&
{\cal I} \biggl \{
-8 \int_1^\infty \stackrel{(4)}{{\cal
I}^{ii}}(t-r\zeta) Q_0(\zeta)d\zeta
+8(1-\ln r)\stackrel{(4)}{{\cal I}^{ii}}(t) \biggr \}
\,,\\
(B_{3.5})_{{\cal C}-{\cal N}} &=&
{\cal I} \biggl \{
+{8 \over 3} {\hat n}^i \int_1^\infty \stackrel{(5)}{{\cal
I}^{ijj}}(t-r\zeta) Q_1(\zeta)d\zeta
+{32 \over 3} \epsilon^{aij}{\hat n}^i \int_1^\infty
\stackrel{(4)}{{\cal J}^{aj}}(t-r\zeta) Q_1(\zeta)d\zeta 
\nonumber \\
&&
+4r \stackrel{(5)}{{\cal I}^{ii}}(t)
-{4 \over 3} {\hat n}^i \stackrel{(5)}{{\cal
I}^{ijj}}(t)
-{16 \over 3} \epsilon^{aij}{\hat n}^i \stackrel{(4)}{{\cal
J}^{aj}}(t) \biggr \}
\,, \\
%(B_4)_{{\cal C}-{\cal N}} &=&
%{\cal I} \biggl \{
%-16 {\hat n}^{<jk>} \int_1^\infty \stackrel{(4)}{{
%M}^{iijk}}(t-r\zeta) Q_2(\zeta)d\zeta
%-{16 \over 3} \int_1^\infty \stackrel{(4)}{{
%M}^{iijj}}(t-r\zeta) Q_0(\zeta)d\zeta \biggr \}
%\,, \\
(B_4)^{ij}_{{\cal C}-{\cal N}} &=&
- 8 {\cal I}
\int_1^\infty \stackrel{(4)}{{\cal
I}^{ij}}(t-r\zeta) Q_0(\zeta)d\zeta \,.
\end{eqnarray}
\label{outerterms}
\end{mathletters}
%label{outerterms}

Using the recursion
relations satisfied by Legendre functions, we can establish
the general formulae:
\begin{eqnarray}
\int_1^\infty X(t-r\zeta) Q_L(\zeta)d\zeta &=& {1 \over {L(L+1)}} X(t-r)
\nonumber \\
&& 
-{1 \over {2L+1}} \int_1^\infty X^\prime (t-r\zeta) 
(Q_{L+1}(\zeta) - Q_{L-1}(\zeta) )d\zeta \,,
\nonumber \\
\int_1^\infty X(t-r\zeta) Q_0(\zeta)d\zeta &=&  X(t-r)
-\int_1^\infty X^\prime(t-r\zeta) (Q_{1}(\zeta) + Q_{0}(\zeta) )d\zeta 
\nonumber \\
&& 
+\int_0^\infty \dot X(t-r-s) \ln (s/2r) ds \,,
\end{eqnarray}
where prime denotes $\partial/\partial \zeta$, $s=r(\zeta -1)$, 
$X$ represents one of the multipole moments of the system
(${\cal I}^{ij}$ and higher), and  we assume
that, in the distant past the system becomes sufficiently
``stationary'' that as $s \to \infty$, $X(t-r-s) \ln s \to 0$.
Since for a binary system that
becomes unbound ($r \to v_0 s$) in the infinite past (because of
gravitational-radiation anti-damping, looking backwards), 
$X$ in the worst 
case is proportional to $(d/dt)^4{{\cal
I}^{ij}} \sim mv^4/r^2 \to mv_0^2/s^2$, then this boundary condition is
satisfied (see \cite{walkerwill2} for a detailed discussion of the past behavior
of binary systems whose evolution includes gravitational radiation
reaction).  Repeated use of these identities allows us to 
convert many of the integrals in Eqs. (\ref{outerterms}) into 
integrals of higher time-derivatives
of the expressions, which are thus of higher PN order, plus residual
terms that cancel many of the non-integral terms in Eqs. (\ref{outerterms}).  
It is also useful to expand the
retarded time $t-r-s$ about $t-s$, and to separate the $\ln r$ 
terms from the $\ln (s/2)$ terms in the integrals, leaving only terms
proportional to $X^{(n)}(t)$ and $\int_0^\infty X^{(n)}(t-s) \ln (s/2)
ds$. 
In the end, the only terms that remain at 3PN and 3.5PN order are
\begin{mathletters}
\begin{eqnarray}
N_{{\cal C}-{\cal N}} &=& {\cal I} \biggl \{
- {16 \over 3} \stackrel{(4)}{{\cal I}^{ii}}(t)
-{8 \over 3} \int_0^\infty \stackrel{(5)}{{\cal I}^{ii}}(t-s) \ln
(s/2) ds \biggr \} +O(\epsilon^5)\,, \\
K^i_{{\cal C}-{\cal N}} &=& O(\epsilon^{9/2}) \,, \\
B^{ij}_{{\cal C}-{\cal N}}&=& O(\epsilon^{4}) \,, \\
B_{{\cal C}-{\cal N}}&=& -8 {\cal I}  \int_0^\infty \stackrel{(5)}{{\cal
I}^{ii}}(t-s) \ln (s/2) ds  +O(\epsilon^5)\,.
\end{eqnarray}
\label{NBresidual}
\end{mathletters}
%label{NBresidual}
As these are purely functions of time, they do not contribute to the
equations of motion through 3.5PN order.
Alternatively, one can show that the terms in Eqs. (\ref{outerterms})
turn out to be purely gauge
terms through 3.5PN order.  In fact, by making the gauge transformation
$h^{\mu\nu} \to h^{\mu\nu} - \xi^{\mu,\nu} - \xi^{\nu,\mu}
+\eta^{\mu\nu} \xi^\alpha_{,\alpha}$ (the linear transformation
suffices to this order), with
\begin{eqnarray}
\xi^0 &=& 
{\cal I} \biggl \{
{4 \over 3} \stackrel{(3)}{{\cal I}^{ii}}(t) 
+{8 \over 3} \int_0^\infty \stackrel{(4)}{{\cal I}^{ii}}(t-s) \ln (s/2) ds
-{2 \over 3} x^{ij} \int_0^\infty \stackrel{(6)}{{\cal I}^{ij}}(t-s)
\ln (s/2) ds
\nonumber \\
&&
\quad + {2 \over 3} r^2 \int_0^\infty \stackrel{(6)}{{\cal I}^{ii}}(t-s)
\ln (s/2) ds
+ {4 \over 45} x^i \int_0^\infty \stackrel{(6)}{{\cal I}^{ijj}}(t-s)
\ln (s/2) ds
\nonumber \\
&&
\quad +{8 \over 9} x^i \epsilon^{ikj} \int_0^\infty \stackrel{(5)}
{{\cal J}^{jk}}(t-s) \ln (s/2) ds
%\nonumber \\
%&&
%+ {16 \over 3}\int_0^\infty \stackrel{(4)} {{M}^{iijj}}(t-s) \ln (s/2) ds
%+{2 \over 5} \int_0^\infty \stackrel{(6)} {{\cal I}^{iijj}}(t-s) \ln (s/2)
%ds
\biggr \} 
\,,\\
\xi^i &=& 
{\cal I} \biggl \{
-4x^j \int_0^\infty \stackrel{(5)}{{\cal I}^{ij}}(t-s) \ln (s/2) ds
+ {4 \over 3}x^i \int_0^\infty \stackrel{(5)}{{\cal I}^{jj}}(t-s) \ln (s/2) ds
\nonumber \\
&&
\quad +{44 \over 45} \int_0^\infty \stackrel{(5)}{{\cal I}^{ijj}}(t-s) \ln
(s/2) ds
-{8 \over 9} \epsilon^{ikj} \int_0^\infty \stackrel{(4)}
{{\cal J}^{jk}}(t-s) \ln (s/2) ds \biggr \} \,,
\end{eqnarray}
we can convert the outer integral contributions to $h^{\alpha\beta}$
in Eq. (\ref{outerterms})
to a form consisting of nothing but a 4PN tail term:
\begin{eqnarray}
(N+B)_{{\cal C}-{\cal N}} &=&
-{16 \over 5} {\cal I} x^{ij} \int_0^\infty \stackrel{(7)\qquad}
{{\cal I}^{<ij>}}(t-s) \ln (s/2) ds  +O(\epsilon^5) \,, \nonumber \\
(K^i)_{{\cal C}-{\cal N}} &=& O(\epsilon^{9/2}) \,, \nonumber \\
(B^{ij})_{{\cal C}-{\cal N}}  &=& O(\epsilon^{4}) \,.
\label{tailfinal4PN}
\end{eqnarray}
%label{tailfinal4PN}
Note that, to this order, $N+B = 2 g_{00}$, and only the gradient
of the term in Eq. (\ref{tailfinal4PN}) 
contributes to the acceleration, hence this 
term can be thought of as a 4PN tail modification of the Newtonian
gravitational potential, or as a 1.5PN modification due to tails of the
2.5PN radiation-reaction potentials.  This result is in complete agreement with
the near-zone tail contribution derived by Blanchet and Damour \cite{bd88} 
using
matched asymptotic expansions within the post-Minkowskian formalism.

\section{Discussion}

We have presented a method for direct integration of the relaxed Einstein
equations in a post-Newtonian expansion, applicable to equations of motion
and gravitational radiation from isolated gravitating systems.  As a
foundation for future work, we presented a solution for the near-zone
gravitational field through 3.5 post-Newtonian order
in terms of Poisson potentials, together with a
prescription for ensuring that no divergent or undefined integrals occur.
In subsequent work, we will apply the near-zone results to the derivation of
equations of motion for binary systems of compact objects through 2.5 PN
order, and including 3.5 PN radiation reaction terms.  Work on the 3PN
contributions to the equations of motion is in progress.  

The results presented here can also be applied to the gravitational
radiation wave-form and energy flux from binary systems to as high as 3PN
order beyond the quadrupole approximation.  It can also be used to discuss
equations of motion and radiation damping of systems containing spinning
bodies, as well as the structure and evolution of fluid bodies.  These will
be the subject of future work.

\acknowledgments
We gratefully acknowledge useful discussions with Luc Blanchet.  This
research was supported in part by the National Science Foundation under
Grant No. PHY 96-00049.  The initial phase of this work
was also supported by Fellowships to CW from the
Fulbright Foundation and the J. S. Guggenheim Foundation.  CW is grateful to
the Observatoire de Paris, Meudon and the Centre National de la Recherche
Scientifique in France, and to the Hebrew University of Jerusalem, for
hospitality and support during a sabbatical year where this work began in
earnest.

\appendix

\section{STF Tensors and their properties}
\label{STF}

Throughout this series of papers, we shall make frequent use of the properties
of symmetric, trace-free (STF) products of unit vectors.
The general
formula for such STF products is
\begin{eqnarray}
\hat n^{<L>} \equiv \sum_{p=0}^{[l/2]} (-1)^p {{(2l-l-2p)!!} \over
{(2l-1)!!}} \left [ \hat n^{L-2P} \delta^P + {\rm sym(q)} \right ] \;,
\label{STFgen}
\end{eqnarray}
%label{STFgen}
where $[l/2]$ denotes the integer just less than or equal to
$l/2$, the capitalized superscripts denote the dimensionality, $l-2p$
or $p$, of
products of $\hat n^i$ or $\delta^{ij}$ respectively,
and ``sym(q)'' denotes all
distinct terms arising from permutations of
indices, where $q=l!/[(2^p p!(l-2p)!]$ is the total number of such
terms (see \cite{bd86,thorne80} for compendia of formulae).  For
convenience, we display the
first several examples explicitly
\begin{mathletters}
\label{STFformulae}
\begin{eqnarray}
\hat n^{<ij>} &=& \hat n^{ij} - {1 \over 3} \delta^{ij} \;, \\
\hat n^{<ijk>} &=& \hat n^{ijk} - {1 \over 5} (\hat n^i \delta^{jk} +\hat
n^j \delta^{ik} + \hat n^k \delta^{ij}) \;, \\
\hat n^{<ijkl>} &=& \hat n^{ijkl} - {1 \over 7}(\hat n^{ij}\delta^{kl}
+ {\rm sym(6)}) +{1 \over
35}(\delta^{ij}\delta^{kl}+\delta^{ik}\delta^{jl}+\delta^{il}\delta^{jk})
\;, \\
\hat n^{<ijklm>} &=& \hat n^{ijklm} - {1 \over 9}(\hat n^{ijk}\delta^{kl}
+ {\rm sym(10)}) +{1 \over
63}(\hat n^i \delta^{jk}\delta^{lm} +{\rm sym(15)})
\;, \\
\hat n^{<ijklmn>} &=& \hat n^{ijklmn} - {1 \over 11}(\hat
n^{ijkl}\delta^{mn}
+ {\rm sym(15)}) +{1 \over
99}(\hat n^{ij}\delta^{kl}\delta^{mn}+ {\rm sym(45)}) \nonumber \\
&&-{1 \over 693} (\delta^{ij}\delta^{kl}\delta^{mn}+ {\rm sym(15)} )
\;.
\end{eqnarray}
\end{mathletters}
%label{STFformulae}
There is a close connection between these STF tensors and spherical
harmonics.  For example, it is straightforward to show that, for any
unit vector $\bf \hat N$, the contraction of $\hat N^L$ with $\hat
n^{<L>}$ is given by
\begin{eqnarray}
\hat N^L \hat n^{<L>} = {{l!} \over {(2l-1)!!}} P_l ({\bf \hat N \cdot
\hat n}) \;,
\label{legendre}
\end{eqnarray}
%label{legendre}
where $P_l$ is a Legendre polynomial.

\section{Cancellation of $\cal R$-dependence between inner and outer
integrals}
\label{cancellation}

Here we demonstrate explicitly the cancellation of $\cal R$-dependent
terms between the inner and outer integrals.  We assume that, at each
iteration step, from just inside the boundary of the near zone out
into the far zone, the source stress-energy tensor
$_{N-1}\Lambda^{\alpha\beta}$ can be decomposed into terms of the form
$f_{B,L}(u){\hat n}^{<L>} r^{-B}$, where $u=t-r$ is retarded time,
and
${\hat n}^{<L>}$ is a STF product of unit radial vectors.  We
calculate the behavior of the inner integral of such a term as the
integration variable approaches $\cal R$ from below with the result
obtained from the outer integral of the same term.  We consider
far-zone and near-zone field points separately.

\subsection{Far-zone field points}

The inner integral is given by Eq. (\ref{genexpand}), with the
multipole moment given by Eq. (\ref{genmoment}).  We want to examine
the behavior of the moment, as $|{\bf x}^\prime| \to {\cal R}$, that
is 
\begin{eqnarray}
M^{\alpha\beta {\bar Q}}(u) &\to& {1 \over {16\pi}} \int^{\cal R}
f_{B,L}(u-r^\prime) {{{{\hat n}}^{\prime <L>}} \over {{r^\prime}^B}}
{x^\prime}^{\bar Q} {r^\prime}^2 dr^\prime d\Omega^\prime 
\nonumber\\
&=& {1 \over 4} \sum_{m=0}^\infty {(-1)^m \over m!} f_{B,L}^{(m)}(u)
G_{B,L,{\bar Q}}^m ({\cal R}) \Delta^{L,{\bar Q}} \,, 
\end{eqnarray}
where the superscript $(m)$ denotes $m$ retarded time derivatives,
and where
\begin{mathletters}
\begin{eqnarray}
\Delta^{L,{\bar Q}} &=& {1 \over {4\pi}} \oint {\hat n}^{<L>} {\hat n}^{\bar
Q}
d\Omega \,, \\
G_{B,L,{\bar Q}}^m ({\cal R}) &=& \int^{\cal R} {r^\prime}^{2+{\bar
q}-B+m}
d{r^\prime} \nonumber \\
&=& \left \{ \begin{array}{ll}
{\cal R}^{3+{\bar q}-B+m}/ ({3+{\bar q}-B+m}) & 3+{\bar q}-B+m \ne 0 \\
\ln {\cal R} & 3+{\bar q}-B+m=0
\end{array} \right . \,.
\end{eqnarray}
\end{mathletters}
Then, from inside ${\cal R}$, 
\begin{equation}
{h^{\alpha\beta}_{\cal N}}_{B,L}  \to 
\sum_{{\bar q}=0}^\infty {{(-1)^{\bar q}} \over {{\bar q}!}}
\sum_{m=0}^\infty {{(-1)^m} \over {m!}} 
\partial_{\bar Q} \left ( {1 \over r} f_{B,L}^{(m)}(u) \right )
G_{B,L,{\bar Q}}^m ({\cal R})\Delta^{L,{\bar Q}} \,.
\end{equation}
It is straightforward to show that
the contraction of $\partial_{\bar Q}$ with $\Delta^{L,{\bar Q}}$ is given by
\begin{equation}
\Delta^{L,{\bar Q}} \partial_{\bar Q} = \left \{ \begin{array}{ll}
0 & {\bar q}<L \\
0 & L+{\bar q} = {\rm odd} \\
{{2^L {\bar q}! (({\bar q}+L)/2)!} \over 
{({\bar q}+L+1)!(({\bar q}-L)/2)!}} |\nabla^2|^{({\bar q}-L)/2} 
\partial_{<L>} & {\bar q} \ge L 
\end{array}
\right . \,.
\end{equation}
Using the fact that
\begin{mathletters}
\begin{eqnarray}
\nabla^2 \left ( {f(u) \over r} \right )&=& {{\ddot f} \over r} \,,\\
\partial_{<L>} \left ( {f(u) \over r} \right )  &=& (-1)^L {\hat n}^{<L>} 
\sum_{k=0}^L
{{(L+k)!} \over {2^k k!(L-k)!}} {{f^{(L-k)}(u)} \over {r^{k+1}}} \,,
\end{eqnarray}
\end{mathletters}
(see eg. \cite{bd86}) 
and redefining summation variables, $q=m+{\bar q}-k$, $j=L-k$, 
we obtain Eqs. (\ref{farinnerlimit}) and
(\ref{farinnerlimit2}).  Evaluating the outer integral for the same term yields
$z$-dependent or $\ln {\cal R}$-dependent terms that are precisely
equal and opposite those of Eq. (\ref{farinnerlimit2}).

\subsection{Near-zone field point}

In the near zone, for $|{\bf x}^\prime| > |{\bf x}|$, Eq.
(\ref{nearnewversion}) together with the specific decomposition of
$\Lambda^{\alpha\beta}$ gives
\begin{equation}
{_N h^{\alpha\beta}_{\cal N}}_{B,L}  \to
{1 \over 4\pi} \sum_{{\bar q}=0}^\infty {{(-1)^{\bar q}} \over {{\bar q}!}}
\sum_{m=0}^\infty {{(-1)^m} \over {m!}} x^{\bar Q} \partial_t^n
\int^{\cal R} f_{B,L} (t-r^\prime)  
{{{\hat n}^{\prime <L>}} \over {{r^\prime}^B}}
\partial_{\bar Q}^\prime ({r^\prime}^{m-1})
 {r^\prime}^2 dr^\prime d\Omega^\prime \,.
\end{equation}
We use the fact that \cite{bd86}
\begin{eqnarray}
\partial_{\bar Q}^\prime {r^\prime}^{m-1} 
&=& \sum_{k=0}^{k_m} {{(2{\bar q}-4k+1)!!} \over
{(2{\bar q}-2k+1)!!}}{m! \over  {(m-2k)!}} \left [ {{(m-2k-1)!!} \over
{(m-2{\bar q}+2k-1)!!}} \right ] 
\nonumber \\
&& \quad \times {{{\bar q}!} \over {2^k k! ({\bar
q}-2k)!}} \delta^K {\hat n}^{\prime <{\bar Q}-2K>} {r^\prime}^{m-{\bar q}-1} \,,
\end{eqnarray}
where $k_m = {\rm lesser \, of} \{[{\bar q}/2],[m/2] \}$, 
$\delta^K$ denotes a product of $K$ Kronecker deltas, the quantity in
square brackets can be evaluated for negative or positive values of
the arguments, and the
expression $\delta^K {\hat n}^{\prime <{\bar Q}-2K>}$
is to be symmetrized on all indices (since the expression
ultimately is to be contracted on $x^{\bar Q}$ no explicit symmetrization is
needed).    
It can then be shown that 
\begin{equation}
{\hat n}^{\bar Q} {1 \over 4\pi} \oint \delta^K {\hat n}^{\prime <{\bar
Q}-2K>} {\hat n}^{\prime <L>} d\Omega^\prime = {L! \over {(2L+1)!!}}
\delta_{L,{\bar q}-2k} {\hat n}^{<L>} \,.
\end{equation}
We then expand
$f(t-r^\prime) = \Sigma_{n=0}^\infty (-1)^n f^{(n)}(t) {r^\prime}^n/n!$, 
integrate over $r^\prime$
toward $\cal R$, rearrange the summations, and define $r=({\bar q}-L)/2$,
and $q=m+n$, and obtain
\begin{equation}
{_Nh^{\alpha\beta}_{\cal N}}_{B,L} \to  \left ( {2 \over r}
\right ) ^{B-2}
{\hat n}^{<L>} \sum_{q=0}^\infty {\cal E}^{{\rm in},q}_{B,L} (z) r^q {{d^q
f_{B,L}(t)} \over {dt^q}} \,,
\label{nearinnerlimit}
\end{equation}
%label{nearinnerlimit}
with
\begin{eqnarray}
{\cal E}^{{\rm in},q}_{B,L} (z) &=& \sum_{r=0}^{[q/2]} \sum_{m=2r}^{q}
{{(-1)^{L+q} (2)^{2+L-B}(L+r)!} \over {(q-m)!(2L+2r+1)!(m-2r)!r!}} 
\left [ {{(m-1-2r)!!} \over {(m-1-2r-2L)!!}} \right ]
\nonumber \\
&&  \qquad  \times \left \{ \begin{array}{ll}
{z^{q-L-2r-B+2}/  (q-L-2r-B+2)} & \quad q-L-2r-B+2 \ne 0 \\
\ln {\cal R}  & \quad q-L-2r-B+2 =0 \,. \end{array}
\right .
\label{nearinnerlimit2}
\end{eqnarray}
%label{nearinnerlimit2}
Here too, evaluating the outer integral for the same term, for each
$B$, $L$ and $q$  yields
$z$-dependent or $\ln {\cal R}$-dependent terms that are precisely
equal and opposite those of Eq. (\ref{nearinnerlimit2}).

\subsection{Source terms with $\ln r$ dependence}

Until now we have assumed that the stress-energy source 
$\Lambda^{\alpha\beta}$ can be decomposed into terms of the form 
$f_{B,L}(u) {\hat n}^{<L>} r^{-B}$.  At sufficiently high PN order, 
tail contributions to  the fields will arise, leading to the
possibility of $\ln r$ dependence in $\Lambda^{\alpha\beta}$.  To
illustrate that cancellation of $\cal R$ dependence occurs in this
event also, we consider source terms of the form 
$f_{B,L}(u^\prime) {\hat n}^{\prime <L>} {r^\prime}^{-B} \ln {r^\prime}$. 
Noting that, from Eq. (\ref{rprime1}), $\ln r^\prime = - \ln [2(\zeta
-y)/r(\zeta^2-1)]$, and incorporating this logarithmic term  into the
outer integral, Eq. 
(\ref{outerfar0}), we obtain
\begin{eqnarray}
{{_Nh^{\alpha\beta (\ln)}_{{\cal C}-{\cal N}}}_{B,L}} &=& - {1 \over 2}
{\hat n}^{<L>} \int_{-1}^1 P_L(y)dy
\int_{\zeta(y)}^\infty \left ({{2(\zeta-y)} \over {r(\zeta^2-1)}} \right
)^{B-2} \ln \left ({{2(\zeta-y)} \over {r(\zeta^2-1)}} \right) 
\nonumber \\
&& \qquad \times f_{B,L} (u -r(\zeta-1)) {d\zeta \over {\zeta -y}}
\nonumber \\
&=& - {\partial \over {\partial B}} {{_Nh^{\alpha\beta}_{{\cal C}-{\cal
N}}}_{B,L}} \,.
\label{houterlog}
\end{eqnarray}
For the inner integral, the only difference which the logarithmic term
makes is in the radial integral,
now given by
\begin{eqnarray}
{G_{B,L,{\bar Q}}^m ({\cal R})}^{(\ln)} &=& \int^{\cal R} {r^\prime}^{2+{\bar
q}-B+m} \ln r^\prime
d{r^\prime} 
\nonumber \\
&=& - {\partial \over {\partial B}} G_{B,L,{\bar Q}}^m ({\cal R}) \,.
\label{houterlog2}
\end{eqnarray} 
Thus, if the original coefficients cancel for all $B$ (and if we can
treat $B$ formally as a continuous parameter), then the coefficients
generated by $\ln r$ terms cancel.

An alternative method is to show directly from the definitions (eg.
Eqs. (\ref{houterlog}) and (\ref{houterlog2}))
that, for both the inner and outer
integrals and for $z<1$ and $z>1$, 
\begin{equation}
{_Nh^{\alpha\beta (\ln)}_{B,L}} = \ln {\cal R} \,
_Nh^{\alpha\beta}_{B,L} - \int_1^z {_Nh^{\alpha\beta}_{B,L}}  
d \bar z / \bar z \,,
\end{equation}
modulo $z$- or $\cal R$-independent terms.  Then, if the $z$-dependent
parts of $_Nh^{\alpha\beta}_{B,L}$ cancel between outer and inner
integrals, so too do the $z$-dependent parts of 
${_Nh^{\alpha\beta (\ln)}_{B,L}}$. 

\section{Boundary Terms}
\label{boundary}

The boundary terms in $h_{\cal N}^{\alpha\beta}$ that arise from integrating
by parts
various integrals over $\cal M$
are given by
\begin{mathletters}
\begin{eqnarray}
N_{\partial {\cal M}} &=&
4 \oint_{\partial {\cal M}} \tau^{0j}(t,{\bf x}^\prime ) d^2S_j^\prime
+{2 \over 3} r^2 \partial_t^2 \oint_{\partial {\cal M}} \tau^{0j}(t,{\bf
x}^\prime ) d^2S_j^\prime
-{4 \over 3} x^i \partial_t \oint_{\partial {\cal M}} \tau^{ij}(t,{\bf
x}^\prime ) d^2S_j^\prime
\nonumber \\
&&
-{4 \over 3} x^i \partial_t^2 \oint_{\partial {\cal M}} \tau^{0j}(t,{\bf
x}^\prime ) {x^\prime}^i d^2S_j^\prime
+{1 \over 30} r^4 \partial_t^4 \oint_{\partial {\cal M}} \tau^{0j}(t,{\bf
x}^\prime ) d^2S_j^\prime
\nonumber \\
&&
-{2 \over 15} r^2 x^i \partial_t^4 \oint_{\partial {\cal M}} \tau^{0j}(t,{\bf
x}^\prime ) {x^\prime}^i d^2S_j^\prime
-{2 \over 15} r^2 x^i \partial_t^3 \oint_{\partial {\cal M}} \tau^{ij}(t,{\bf
x}^\prime ) d^2S_j^\prime \,, \\
K^i_{\partial {\cal M}} &=&
4 \oint_{\partial {\cal M}} \tau^{ij}(t,{\bf x}^\prime ) d^2S_j^\prime
+{2 \over 3} r^2 \partial_t^2 \oint_{\partial {\cal M}} \tau^{ij}(t,{\bf
x}^\prime ) d^2S_j^\prime
+{2 \over 3} x^k \partial_t^3 \oint_{\partial {\cal M}} \tau^{0j}(t,{\bf
x}^\prime ) {x^\prime}^{ik} d^2S_j^\prime
\nonumber \\
&&
-{4 \over 3} x^k \partial_t^2 \oint_{\partial {\cal M}} \tau^{j[i}(t,{\bf
x}^\prime ) {x^\prime}^{k]} d^2S_j^\prime
-{2 \over 9} \partial_t^3 \oint_{\partial {\cal M}} \tau^{0j}(t,{\bf
x}^\prime ) {r^\prime}^2 {x^\prime}^i d^2S_j^\prime
\,, \\
B^{ij}_{\partial {\cal M}} &=&
-4 \partial_t \oint_{\partial {\cal M}} \tau^{k(i}(t,{\bf
x}^\prime ) {x^\prime}^{j)} d^2S_k^\prime
-2 \partial_t^2 \oint_{\partial {\cal M}} \tau^{0k}(t,{\bf
x}^\prime ) {x^\prime}^{ij} d^2S_k^\prime
\nonumber \\
&&
-{2 \over 3} r^2 \partial_t^3 \oint_{\partial {\cal M}} \tau^{k(i}(t,{\bf
x}^\prime ) {x^\prime}^{j)} d^2S_k^\prime
-{1 \over 3}  r^2 \partial_t^4 \oint_{\partial {\cal M}} \tau^{0k}(t,{\bf
x}^\prime ) {x^\prime}^{ij} d^2S_k^\prime
\nonumber \\
&&
+{2 \over 3} x^l \partial_t^3 \oint_{\partial {\cal M}} \tau^{k(i}(t,{\bf
x}^\prime ) {x^\prime}^{jl)} d^2S_k^\prime
+{2 \over 9} x^l \partial_t^4 \oint_{\partial {\cal M}} \tau^{0k}(t,{\bf
x}^\prime ) {x^\prime}^{ijl} d^2S_k^\prime
\nonumber \\
&&
+{8 \over 9} x^l \partial_t^3 \oint_{\partial {\cal M}} (\tau^{k[i}(t,{\bf
x}^\prime ) {x^\prime}^{l]j} + \tau^{k[j}(t,{\bf
x}^\prime ) {x^\prime}^{l]i})d^2S_k^\prime
\nonumber \\
&&
-{1 \over 18} \partial_t^3 \oint_{\partial {\cal M}} [ \tau^{lk}
(t,{\bf x}^\prime )({r^\prime}^2 {x^\prime}^{ij})_{,l} + 
{\dot \tau}^{0k}(t,{\bf x}^\prime ) {r^\prime}^2
{x^\prime}^{ij} ] d^2S_k^\prime
\nonumber \\
&&
+{1 \over 3}\partial_t^3 \oint_{\partial {\cal M}} (\tau^{k[l}(t,{\bf
x}^\prime ) {x^\prime}^{i]jl} + \tau^{k[l}(t,{\bf
x}^\prime ) {x^\prime}^{j]il})d^2S_k^\prime 
\nonumber \\
&&
-{1 \over 30} r^4 \partial_t^5 \oint_{\partial {\cal M}}
\tau^{k(i}(t,{\bf
x}^\prime ) {x^\prime}^{j)} d^2S_k^\prime
-{1 \over 60}  r^4 \partial_t^6 \oint_{\partial {\cal M}}
\tau^{0k}(t,{\bf
x}^\prime ) {x^\prime}^{ij} d^2S_k^\prime
\nonumber \\
&&
+{1 \over 15} r^2 x^l \partial_t^5 \oint_{\partial {\cal M}}
\tau^{k(i}(t,{\bf
x}^\prime ) {x^\prime}^{jl)} d^2S_k^\prime
+{1 \over 45} r^2 x^l \partial_t^6 \oint_{\partial {\cal M}}
\tau^{0k}(t,{\bf
x}^\prime ) {x^\prime}^{ijl} d^2S_k^\prime
\nonumber \\
&&
+{4 \over 45} r^2 x^l \partial_t^5 \oint_{\partial {\cal M}}
(\tau^{k[i}(t,{\bf
x}^\prime ) {x^\prime}^{l]j} + \tau^{k[j}(t,{\bf
x}^\prime ) {x^\prime}^{l]i})d^2S_k^\prime \,.
\end{eqnarray}
\label{surfaceterms}
\end{mathletters}
%label{surfaceterms}

\section{Properties of Poisson Potentials}
\label{poisson}

Here we list some useful properties of Poisson potentials and
superpotentials, given by Eqs. (\ref{definepoisson}).  These rely upon the
general result, which can be obtained by integration by parts,
\begin{equation}
P(\nabla^2 g) = -g + {\cal B}_P(g) \,,
\end{equation}
where ${\cal B}_P(g)$ denotes the boundary term, given by
\begin{equation}
{\cal B}_P(g) \equiv {1 \over {4\pi}} \oint_{\partial {\cal M}} \biggl
[
{{g(t,{\bf x}^\prime)} \over {|{\bf x}-{\bf x}^\prime |}} \partial_r^\prime
\ln (g(t,{\bf x}^\prime)|{\bf x}-{\bf x}^\prime |) \biggr ]_{r^\prime
={\cal R}} {\cal R}^2 d\Omega^\prime \,.
\label{Pformulae1}
\end{equation}
%label{Pformulae1}
The boundary terms must be carefully evaluated case by case to
determine if any $\cal R$-independent terms survive.  All $\cal
R$-{\it dependent} terms can be discarded.  Some useful formulae that
result from this include:
\begin{mathletters}
\begin{eqnarray}
P(|\nabla g|^2) &=& -{1 \over 2} \{ g^2 + 2 P(g\nabla^2 g) - {\cal
B}_P(g^2) \} \,, \\
P(\nabla g \cdot \nabla f) &=& -{1 \over 2} \{ fg +  P(f\nabla^2 g) +
 P(g\nabla^2 f) - {\cal B}_P(fg) \} \,, \\
P(f|\nabla U|^2) &=& -{1 \over 2} \{ fU^2 + P(U^2 \nabla^2 f)
-2\Sigma(fU) + 4P(U\nabla U \cdot \nabla f) - {\cal B}_P(fU^2) \} \,.
\end{eqnarray}
\label{Pformulae2}
\end{mathletters}
%label{Pformulae2}
In many specific cases, the boundary terms can be dropped:
\begin{mathletters}
\begin{eqnarray}
P(U) &=& -{1 \over 2}X  \,, \\
P(X) &=& - {1 \over 12}Y \,, \\
P(|\nabla U|^2) &=& - {1 \over 2}U^2 + \Phi_2 \,,\\
P(x^i U^{,jk\dots}) &=& -{1 \over 2}x^i X^{,jk\dots} + {1 \over
12}Y^{,ijk\dots} \,,\\
P(r^2 U^{,ij}) &=& -{1 \over 2}r^2X^{,ij} -{1 \over 12}Y^{,ij} +
{1 \over 6} x^k Y^{,ijk} \,, 
\end{eqnarray}
\label{Pformulae3}
\end{mathletters}
%label{Pformulae3}
while in others, there are contributions from the boundary terms.  For
example, in the 2PN potential $P(\nabla U \cdot \nabla {\ddot X})$,
the boundary term yields the term ${1 \over 2} \int_{\cal M} \sigma(t,{\bf x})
d^3x \partial_t^2 \int_{\cal M} {\sigma}(t,{\bf y}) d^3y$.  Using Eq.
(\ref{massidentity}), we obtain, to the necessary order,
\begin{equation}
P(\nabla U \cdot \nabla {\ddot X}) = -{1 \over 2} \{ U {\ddot X} -
\Sigma(\ddot X) + 2G_2 - {1 \over 2}{\cal I}\stackrel{(4)\quad}{{\cal
I}^{ii}(t)} \}  + O(\epsilon^5)  \,. \\ 
\end{equation}
Similarly, we find for the 3PN potential,
\begin{equation}
P(\nabla U \cdot \nabla \stackrel{(4)}{Y}) = -{1 \over 2} \{ U \stackrel{(4)}{Y}
- \Sigma(\stackrel{(4)}{Y}) + 12P(U \stackrel{(4)}{X}) 
- 2{\cal I}\stackrel{(4)\quad}{{\cal I}^{ii}(t)} \}+ O(\epsilon^5)  \,. \\
\end{equation}
For the Poisson Superpotential $S(f)$, we have
\begin{equation}
S(\nabla^2 g) = 2P(g) + {\cal B}_S(g) \,,
\end{equation}
where 
\begin{equation}
{\cal B}_S(g) \equiv {1 \over {4\pi}} \oint_{\partial {\cal M}} \biggl
[
{g(t,{\bf x}^\prime)} {|{\bf x}-{\bf x}^\prime |}
\partial_r^\prime
\ln \left ( {{g(t,{\bf x}^\prime)} \over {|{\bf x}-{\bf x}^\prime |}} \right
) \biggr ]_{r^\prime
={\cal R}} {\cal R}^2 d\Omega^\prime \,.
\end{equation}
Thus, for example, in the superpotential $(\partial/\partial t)^2
\int_{\cal M} \tau^{00} |{\bf x}-{\bf x}^\prime | d^3x^\prime$, we
find the term
\begin{eqnarray}
\ddot S(\nabla^2 U^2) &=& 2\ddot P(U^2) -3 (d/dt)^2 (\int_{\cal M} 
\sigma d^3x)^2 +O(\epsilon^5) \nonumber \\
&=& 4G_1 +4G_2 - 3{\cal I}\stackrel{(4)\quad}{{\cal I}^{ii}(t)}
+O(\epsilon^5)  \,.
\end{eqnarray}
Other useful identities include
\begin{mathletters}
\begin{eqnarray}
\Sigma(x^i) &=& x^i U -X^{,i} \,, \\
\Sigma(x^{ij}) &=& {1 \over 3} Y^{,ij} -\delta^{ij} X + x^{ij}U
-2x^{(i} X^{,j)} \,.
\end{eqnarray}
\end{mathletters}

\end{document}